\documentclass[a4paper,11pt]{article}
\usepackage{newpxtext}

\usepackage[a4paper,left=1in,right=1in,top=1in,bottom=1in,marginparwidth=1.75cm]{geometry}
\usepackage{soul}
\usepackage[american]{babel}
\usepackage{csquotes}
\usepackage[authoryear,round]{natbib}

\usepackage{amsmath}
\usepackage{amsfonts}
\usepackage{amssymb}
\usepackage{amsthm}
\usepackage{newpxmath} 

\usepackage{chngcntr}
\usepackage{graphicx}
\usepackage{subcaption}
\usepackage{fancyhdr}
\usepackage{tabularx}
\usepackage{geometry}
\usepackage{setspace}
\usepackage[right]{eurosym}
\usepackage[printonlyused]{acronym}
\usepackage{floatflt}
\usepackage[capposition=top]{floatrow}
\usepackage{colortbl}
\usepackage{paralist}
\usepackage{array}
\usepackage{titlesec}
\usepackage{parskip}
\usepackage[right]{eurosym}
\usepackage{longtable}
\usepackage[colorlinks=true,linkcolor=black,urlcolor=blue,citecolor=black,pdfpagelabels=true]{hyperref}
\usepackage{mathtools}
\usepackage{booktabs}
\usepackage{caption}
\allowdisplaybreaks

\usepackage[flushleft]{threeparttable}
\usepackage{hyperref}
\usepackage{dsfont}
\usepackage{enumitem}
\usepackage[toc,page]{appendix}
\usepackage{rotating}
\usepackage{bbold}
\usepackage{scrextend}
\usepackage{xcolor}
\usepackage{algorithm2e}

\usepackage{listings}
\makeatletter
\def\l@lstlisting#1#2{\@dottedtocline{1}{0em}{1em}{\hspace{1,5em} Lst. #1}{#2}}
\makeatother

\newcommand\posscite[1]{\citeauthor{#1}'s (\citeyear{#1})}

\numberwithin{equation}{section}

\theoremstyle{plain}
\newtheorem{assumption}{Assumption}

\newtheorem{theorem}{Theorem}

\newtheorem{proposition}{Proposition}
\newtheorem{lemma}{Lemma}[section]

\theoremstyle{remark}
\newtheorem{remark}{Remark}
\theoremstyle{plain}

\begin{document} \singlespacing \pagestyle{plain}
	
	\begin{center}
		{\Large Informative Local Average Treatment Effects Using Propensity Scores}
		
		{\large \vspace{0.7 cm}}
		
		{\large Martin Huber* and Ying-Ying Lee$^\dag$}\smallskip
		
		{\small {*University of Fribourg, Dept.\ of Economics}\\
			{$^\dag$University of California, Irvine, Dept.\ of Economics}\\
			\smallskip {\ \\[0pt]
				\textbf{}}}\smallskip
	\end{center}
	
	\noindent \textbf{Abstract:} {\small 
In instrumental variable (IV) models, comparisons of conditional local average treatment effects across covariates can mix effect heterogeneity in observed covariates with changes in the complier population, that is, those whose treatment responds to the instrument. Compliers at different covariate values may differ in unobserved characteristics that may also drive effect heterogeneity. We propose informative conditional local average treatment effects (ICLATEs), which condition on a subset of covariates of interest and on the treatment propensity scores under both values of a binary instrument. Fixing these scores holds constant the range of unobservables that drive treatment selection and thus compliance, while the composition of the covariates not conditioned on may change. Under conditional IV assumptions and suitable support conditions, we establish nonparametric identification and characterize the relationship of the ICLATEs to marginal treatment effects. We distinguish effect homogeneity in the unobservables driving treatment selection from effect homogeneity across excluded covariates. We clarify their roles in extrapolating complier effects to population effects and show that, jointly, they yield a testable implication even with a single binary instrument. For estimation, we combine cross-fitted machine learning of the nuisance functions with pooled local linear regression of doubly robust scores. Under undersmoothing and regularity conditions, we establish pointwise asymptotic normality, which a simulation study supports. As an empirical illustration, we apply the method to the Job Corps program to assess how the earnings effects of training participation vary with age.
	\par}\bigskip
    
	
	{\footnotesize
		\textbf{Acknowledgments:} We are deeply indebted to Michael Lechner (University of St.\ Gallen) and Michael Zimmert (Swiss National Bank), who proposed the original idea for this paper to us and contributed to the initial identification results. \textbf{Addresses for correspondence:} Martin Huber, Chair of Applied Econometrics - Evaluation of Public Policies, University of Fribourg, Bd.\ de P\'{e}rolles 90, 1700 Fribourg, Switzerland, \href{mailto:martin.huber@unifr.ch}{martin.huber@unifr.ch}. Ying-Ying Lee, Department of Economics, University of California Irvine, Irvine, CA 92697, U.S.A.,  \href{mailto:yingying.lee@uci.edu}{yingying.lee@uci.edu}}.
	
	\thispagestyle{empty}\pagebreak {\small \renewcommand{\thefootnote}{%
			\arabic{footnote}} \setcounter{footnote}{0} \pagebreak
		\setcounter{footnote}{0} \pagebreak \setcounter{page}{1} }
	
	\clearpage
	\doublespacing
	\section{Introduction}
	
		In causal inference based on instrumental variables (IV), the notion of treatment effect heterogeneity is used in at least two distinct senses. One sense concerns heterogeneity that originates in the treatment selection mechanism itself: in the framework of \citet{Imbens_Angrist_1994}, a binary instrument nonparametrically identifies the average effect among compliers, i.e.\ among those units whose treatment status responds to the instrument. The local average treatment effect (LATE) is therefore defined by the instrument at hand, and different instruments generally identify effects for different subpopulations. The other sense concerns effects conditional on observed covariates, as studied for instance by \citet{Abrevaya_Hsu_Lieli_2015}, \citet{Lee_Okui_Whang_2016}, \citet{Wager_Athey_2018}, \citet{Lechner_2018}, \citet{Zimmert_Lechner_2019}, \citet{Fan_Hsu_Lieli_Zhang_2019}, and \citet{Semenova_Chernozhukov_2021}, who target conditional average treatment effects under selection on observables. A number of contributions combine the two notions and consider complier effects conditional on covariates, building on \citet{Abadie_2003} and \citet{Frolich_2007}; see \citet{Ogburn_Rotnitzky_Robins_2015} and \citet{Athey_Tibshirani_Wager_2019}. The resulting conditional LATE (CLATE) is, we argue, only partially informative, because the complier population it refers to changes with the covariates it conditions on. Apparent heterogeneity may therefore reflect the covariates themselves, unobservables that differ across complier populations, or both. More flexible estimation of the CLATE does not resolve this, because the difficulty is one of interpretation rather than of estimation.
	
	To resolve this ambiguity in effect heterogeneity, we propose a family of causal parameters that condition not only on a subset of covariates of interest, but also on the probabilities of being treated under either value of the instrument. The conventional IV assumptions imply a threshold-crossing representation of the treatment decision \citep{Vytlacil_2002}, 
   under which these two treatment propensity scores are exactly the quantiles of the distribution of the unobserved term that delimit the complier population, so that holding them fixed holds the range of unobservables fixed.
    We call the resulting parameters informative conditional local average treatment effects (ICLATEs) and consider three of them. The first conditions on both the subset of covariates of interest and the pair of propensity scores,
    so that comparisons across those covariates no longer mix effect heterogeneity in them with shifts in the composition of unobservables.
    
The second averages over the covariates of interest while still holding the pair of propensity scores (and thus, the unobservables driving compliance) fixed. The third averages over the propensity scores and corresponds to an average complier effect conditional on the covariates of interest. The first two parameters hold that range of unobservables fixed, the third does not. And none of them holds fixed the identity of the compliers or the distribution of the covariates outside the conditioning set, which may still change with the covariates of interest. Attributing variation in the effect across the covariates of interest to those covariates alone therefore requires further restrictions. We establish nonparametric identification of all three parameters, relate them to the marginal treatment effect, and derive an estimator and its asymptotic distribution. 
	
	A running example may make the issue tackled by our approach more concrete. Suppose the researcher studies the wage returns to attending college and uses, conditional on observed covariates, an indicator for living close to a college as an instrument for attendance, as in \citet{Card_1995}. By the result of \citet{Imbens_Angrist_1994}, the instrument identifies the average effect among those who attend college when living close to one and do not attend when living far from one. The identifying assumptions permit selection into treatment to depend on unobserved characteristics, such as the quality of the high school attended. Suppose we are interested in how the return varies with parental earnings. The CLATE, defined as the complier effect conditional on parental earnings, is identified. But the complier population itself changes with parental earnings, because the first-stage effect of proximity on attendance generally differs across parental earnings levels. 
    A comparison of CLATEs across two parental earnings levels therefore mixes a covariate effect with a shift in the range of unobserved high school quality that defines compliance, and is for that reason difficult to interpret. Conditioning in addition on the two propensity scores removes the second channel, motivating the consideration of the ICLATE approach.

	Our contribution is threefold. First, we show that all three suggested ICLATE parameters are nonparametrically identified under the conditional LATE assumptions of \citet{Abadie_2003} and \citet{Frolich_2007}, together with a specific support condition, as the ratio of two conditional expectations of doubly robust score functions in the sense of \citet{Robins_Rotnitzky_Zhao_1994}. No effect homogeneity restriction is needed for these identification results. The identification argument removes the latent compliance indicator from the conditioning set by a Bayes-rule argument that exploits the fact that both the covariates of interest and the propensity scores are measurable functions of the full covariate vector. 
    
    Second, we characterize the relation of the parameters to the marginal treatment effect of \citet{Heckman_Vytlacil_1999,Heckman_Vytlacil_2005}, the average treatment effect conditional on (all) covariates and the unobservable driving selection into treatment: each is an average of marginal treatment effects over an interval of the normalized unobservable, and the three differ in whether that interval is held fixed and in how covariate cells are weighted.
	
	Third, we discuss two effect homogeneity restrictions in the context of the CLATE and ICLATE. One restricts the dependence of the effect on the unobservable while leaving heterogeneity across covariates unrestricted, and it is what permits extrapolation from the complier effect to the population average effect. The other rules out effect heterogeneity across the covariates excluded from the conditioning set while leaving heterogeneity in the unobservable unrestricted. The joint satisfaction of both homogeneity restrictions implies  that the first ICLATE parameter does not vary with the propensity scores at a fixed value of the covariates of interest. This provides a testable implication of the joint homogeneity restrictions. Under the maintained IV assumptions, maintaining either restriction allows rejection to be attributed to the other. Furthermore, if the homogeneity restriction on the excluded covariates is imposed a priori, it in addition follows that the first ICLATE parameter becomes a difference quotient of a single univariate function of the unobservable. 
    Under the additional support and continuity conditions
presented in Theorem~\ref{thm:mtecurve} below, this identifies the marginal
treatment effect curve at the level of the covariates
of interest on the interior of the supported interval.

 We develop feasible pointwise inference when the conditioning
propensity scores are estimated. The estimator combines cross-fitted
nuisance estimation with pooled local linear regression of doubly robust
scores, building on
\citet{Chernozhukov_Chetverikov_Demirer_Duflo_Hansen_Newey_2017}
and the conditional-effect methods of
\citet{Zimmert_Lechner_2019} and \citet{Fan_Hsu_Lieli_Zhang_2019}.
The first step can use parametric, nonparametric, or machine-learning
estimators satisfying the stated conditions.
A distinctive complication is that the treatment propensity functions
enter both the scores (estimated response) and the conditioning index (generated regressors).
With continuous conditioning variables, the local linear
estimator converges at a nonparametric rate. We control the error in the estimated nuisance functions
through partial-$L_2$ consistency and product-rate conditions analogous to
those in \citet{CL}. 
The error from estimating the propensity scores that
enter the conditioning set is controlled separately
through uniform first-stage accuracy and local continuity
of the index-bias projection.
Here, population {\it index bias} is the difference between the conditional mean of the outcome score given the full covariate vector and its conditional mean given the two treatment propensity scores and the covariates of interest.
The continuity condition permits nonzero index bias
and relates to the generated-regressor analyses of
\citet{Mammen_Rothe_Schienle_2012,MRS2016,LeePMGR}.
Under these conditions and undersmoothing, we establish asymptotic normality
with the same first-order expansion and limiting variance as an estimator
that uses the true nuisance functions, and provide a consistent variance
estimator. 
Cross-fitting permits a pointwise argument without entropy
restrictions or derivatives of fitted nuisance functions.

    Our simulation study compares designs with zero and
nonzero population index bias, while keeping the
true value of the target parameter the same.  
    Bias is small relative to sampling variation, standard
deviations decline with sample size, and confidence
interval coverage is generally at or above the nominal
level.
     As an empirical illustration, we apply the proposed method to the Job Corps training program \citep{Schochet_Burghardt_McConnell_2008} to assess the effect of training participation on the weekly earnings of disadvantaged youth. Holding the range of unobservables fixed, we vary age to examine how the ICLATEs change across the age distribution. We find statistically significant positive effects at age 20. For the same complier profile at younger ages, however, the estimated effects are closer to zero and not statistically significant, although the standard errors are relatively large. This makes it difficult to draw precise conclusions about how the effects evolve with age. 
	
	The paper relates to several strands of the literature. \citet{Ogburn_Rotnitzky_Robins_2015} and 
    \citet{Ogburn_Rotnitzky_Robins_2015} and
\citet{Athey_Tibshirani_Wager_2019} study complier effects across
covariates. \citet{Zhao_Ding_Li_2026} characterize when two-stage
least squares with treatment--covariate interactions recovers the
linear projection of treatment effects on covariates among compliers.
Our first ICLATE allows us to compare complier effects across values
of the covariates of interest while holding fixed the two treatment
propensity scores and hence the range of unobserved resistance to
treatment that defines compliance. \citet{Angrist_FernandezVal_2013} and \citet{Aronow_Carnegie_2013} extrapolate the LATE to broader populations under the assumption that, conditional on covariates, average effects do not differ across compliance types. \citet{Wang_TchetgenTchetgen_2018} show that the conditional LATE identifies the conditional average treatment effect if either the treatment effect or the first-stage effect of the instrument is homogeneous in the unobservables that affect the outcome. 
\citet{Brinch_Mogstad_Wiswall_2017} study identification
of marginal treatment effects with discrete instruments
under additional functional restrictions.
Our baseline target is instead an interval-average effect
conditional on the two treatment propensity scores,
which can be identified with a binary instrument without
recovering the entire marginal treatment effect curve.  
    \citet{Mogstad_Torgovitsky_2024} survey this literature.
	
	Because the conditioning set contains estimated propensity scores, the paper also connects to the literature on generated regressors, from \citet{Hahn_Ridder_2013} and \citet{Mammen_Rothe_Schienle_2012} to the locally robust treatment of machine learning generated regressors in \citet{Escanciano_PerezIzquierdo_2026}. Our parameters are close in spirit to the balanced group average treatment effect of \citet{Bearth_Lechner_2025}, who hold the distribution of a set of a priori determined covariates fixed when comparing group average effects under selection on observables; the object we hold fixed is instead the range of the unobservable that defines compliance.
	
	The remainder of the paper is organized as follows. Section~\ref{sec:setting} sets out the IV framework and defines the parameters of interest. Section~\ref{sec:ident} presents the identification results. Section~\ref{sec:relation} relates the parameters to the marginal treatment effect. Section~\ref{sec:extrapolation} discusses extrapolation to average treatment effects and the associated testable implications. Section~\ref{sec:estimation} proposes an estimation strategy and Section~\ref{sec:asymptotics} derives its asymptotic properties. Section~\ref{sec:simulation} presents a simulation study, Section~\ref{sec:application} an application to the Job Corps program, and Section~\ref{con} concludes.

	\section{IV setting and notation}\label{sec:setting}
	
	Let $Y$ denote the outcome variable and $D$ the binary treatment of interest. To describe causal effects we use \posscite{Rubin_1974} potential outcome notation, so that $Y(d)$ is the outcome that would be observed under treatment state $D=d$. Let $Z$ be a binary instrument and write the potential treatment states as $D(z)$. Finally, let $X$ denote a vector of observed covariates with support $\mathcal{X}$. Following \citet{Heckman_Vytlacil_1999}, \citet{Heckman_Vytlacil_2005}, and \citet{Heckman_Urzua_Vytlacil_2006}, we let $V$ be a scalar index of unobserved characteristics and assume that treatment assignment follows a threshold-crossing model, in which the treatment is an additively separable function of the observables $(Z,X)$ on the one hand and of the unobserved term $V$ on the other, and equals one whenever the former exceeds the latter:
	\begin{eqnarray}\label{treatmodel}
		D=I\{\psi(Z,X)\geq V\},
	\end{eqnarray}
	where $I\{\cdot\}$ is the indicator function and $\psi$ is a general, nontrivial function. All compliance types $T$ can be described in terms of $V$ and $\psi(Z,X)$. Compliers ($T=c$), defined by $D(1)=1$ and $D(0)=0$, are those units for which, conditional on $X$, the unobserved term lies in the interval $\psi(0,X)< V \leq \psi(1,X)$. Analogously, never takers ($T=n$) satisfy $V> \psi(1,X)$ and always takers ($T=a$) satisfy $V\leq \psi(0,X)$.
	
	Rather than working with $V$ directly, it is convenient to reparametrize the model in terms of the conditional cumulative distribution function (cdf) of $V$ given $(Z,X)$, denoted by $F_{V|Z,X}$. The random variable $F_{V|Z,X}(V)$ is by construction uniformly distributed on the unit interval. As discussed for instance in \citet{Heckman_Vytlacil_2005}, equation \eqref{treatmodel} may therefore be rewritten without loss of generality as
	\begin{eqnarray}\label{eq:FirstStage}
		D=I\{F_{V|Z,X}(\psi(Z,X))\geq F_{V|Z,X}(V)\},
	\end{eqnarray}
	where $F_{V|Z,X}(\psi(Z,X))$ denotes the conditional cdf of $V$ evaluated at $\psi(Z,X)$. By the definition of a cdf and by \eqref{treatmodel}, $F_{V|Z,X}(\psi(Z,X))=\Pr(V\leq \psi(Z,X)|Z,X)=\Pr(D=1|Z,X)$, which is the treatment propensity score given $Z$ and $X$, henceforth denoted by $\pi(Z,X)$. Writing $\bar{V}=F_{V|Z,X}(V)$ for the normalized unobservable, the compliance types are equivalently characterized by $\pi(0,X)<\bar{V}\leq \pi(1,X)$ for compliers, $\bar{V}>\pi(1,X)$ for never takers, and $\bar{V}\leq\pi(0,X)$ for always takers. Under Assumption~\ref{ass:independence_cov} below, $\bar{V}$ is uniformly distributed on $[0,1]$ and independent of $(Z,X)$, a property that we exploit repeatedly. Accordingly, the LATE conditional on $X$ (CLATE) may be written as
	\begin{eqnarray}\label{CLATE}
		\text{CLATE}=\Delta(x,c)&=&E\left[Y(1)-Y(0)\big|X=x, T=c \right]\notag\\
		&=&E\left[Y(1)-Y(0)\big|X=x,\ \psi(0,x)<V\leq \psi(1,x)\right]\notag\\
		&=&E\left[Y(1)-Y(0)\big|X=x,\ \pi(0,x)<\bar{V}\leq \pi(1,x)\right].
	\end{eqnarray}
	
	To investigate effect heterogeneity across observed characteristics, we introduce further notation. Let $\tilde{X}\subseteq X$ denote a subvector of $X$ with support $\tilde{\mathcal{X}}$. Our aim might be to assess effect heterogeneity with respect to $\tilde{X}$ while holding the complier population fixed in terms of the distribution of $V$. To this end, let $\pi'$ denote a specific value of the propensity score under $Z=0$ and $\pi''$ a specific value of the propensity score under $Z=1$, with $\pi',\pi''\in[0,1]$, and collect the two propensity scores and the two evaluation points in $Q=\left(\pi(0,X),\pi(1,X)\right)^\top$ and $q=(\pi',\pi'')^\top$, so that the event $\{Q=q\}$ is shorthand for $\{\pi(0,X)=\pi',\,\pi(1,X)=\pi''\}$. Our first parameter of interest is the LATE conditional on $\tilde{X}=\tilde{x}$ and on $Q=q$:
	\begin{eqnarray}\label{firsteff}
		\text{ICLATE}_1=\Delta(\tilde{x},c,\pi',\pi'')=E\left[Y(1)-Y(0)|\tilde{X}=\tilde{x},T=c,Q=q\right].
	\end{eqnarray}
	The pair $(\pi',\pi'')$ delimits the range of the distribution of $\bar V$ over which compliers with $\tilde{X}=\tilde{x}$ are defined. Varying $\tilde{x}$ at fixed $(\pi',\pi'')$ therefore compares complier groups defined over the same range of unobservables, which permits the assessment of effect heterogeneity with respect to observables alone. This is exactly what the CLATE in general does not do.
	
	However, it is important to note that such a comparison does not keep the remaining covariates (those not included in $\tilde{X}$) constant. Conditioning on $(\pi',\pi'')$ restricts $X\setminus\tilde{X}$ to a level set that shifts both with $\tilde{x}$ and with $(\pi',\pi'')$. Effect heterogeneity across $\tilde{x}$ at fixed $(\pi',\pi'')$ therefore reflects effect heterogeneity with respect to $\tilde{x}$ and to a changing composition of $X\setminus\tilde{X}$. Effect heterogeneity across $(\pi',\pi'')$ at fixed $\tilde{x}$ reflects effect heterogeneity with respect to unobservable $V$, and again to a changing composition of $X\setminus\tilde{X}$. Section~\ref{sec:extrapolation} returns to how the two channels can be separated based on specific assumptions.
	
	Averaging $\Delta(\tilde{x},c,\pi',\pi'')$ over $\tilde{X}$ gives our second parameter, the LATE conditional on the propensity scores alone:
	\begin{eqnarray}\label{secondeff}
		\text{ICLATE}_2&=&\Delta(c,\pi',\pi'')=E\left[\Delta(\tilde{X},c,\pi',\pi'')|T=c,Q=q\right]\notag\\
		&=&E\left[Y(1)-Y(0)|T=c,Q=q\right].
	\end{eqnarray}
	Averaging instead over the propensity scores yields our third parameter, the LATE conditional on $\tilde{X}=\tilde{x}$. It corresponds to the average effect across complier groups that share the same value of $\tilde{X}$ but differ in $\pi(0,X)$ and $\pi(1,X)$, and hence in the ranges of $V$ over which they are defined:
	\begin{eqnarray}\label{thirdeff}
		\text{ICLATE}_3=\Delta(\tilde{x}, c)&=&E\left[\Delta(\tilde{x},c,\pi(0,X),\pi(1,X))\big|T=c,\tilde{X}=\tilde{x}\right]\notag\\
		&=&E\left[Y(1)-Y(0)\big|T=c,\tilde{X}=\tilde{x}\right].
	\end{eqnarray}
	$\text{ICLATE}_3$ is the parameter most closely related to those considered in the previous literature on complier effects across covariates, see \citet{Ogburn_Rotnitzky_Robins_2015} and \citet{Athey_Tibshirani_Wager_2019}.
	
	The three parameters differ in what they hold fixed, and hence in what they support. $\text{ICLATE}_1$ is the most granular effect holding both $\tilde{X}$ and the range of unobservables fixed. $\text{ICLATE}_2$ holds the range of unobservables (defining compliance) fixed but averages over $\tilde{X}$.  Comparing it across $(\pi',\pi'')$ permits assessing effect heterogeneity across different complier groups, with the caveat that the composition of the entire covariate vector, rather than only of $X\setminus\tilde{X}$, moves along. $\text{ICLATE}_3$ exclusively holds a subset of covariates constant. Assessing effect heterogeneity across  $\tilde{x}$ neither keeps complier groups, nor covariates constant, which must be borne in mind when interpreting the results. Finally, it is worth noting that $\text{ICLATE}_1$ and $\text{ICLATE}_2$ coincide whenever the effect is homogeneous in the subset of covariates $\tilde{X}$, which is testable. Furthermore, $\text{ICLATE}_1$ and $\text{ICLATE}_3$ coincide whenever the effect is homogeneous in $V$ and in $X\setminus\tilde{X}$. The testability of this homogeneity condition is discussed in Section~\ref{sec:extrapolation}.
	
	\section{Identification}\label{sec:ident}
	
	This section states the assumptions under which the CLATE and, building on it, the ICLATE parameters are identified. We first impose the conditional LATE assumptions of \citet{Abadie_2003} and \citet{Frolich_2007}, which are versions of the LATE assumptions of \citet{Imbens_Angrist_1994} and \citet{Angrist+96} that hold conditional on $X$.
	
	\begin{assumption}[Conditional IV validity]\label{ass:independence_cov}
		For all $x\in\mathcal{X}$ and all $z,z',d\in\{0,1\}$:
		$(i)$ $Z \bot \left(D(z), Y(z',d)\right)|X=x$, and $(ii)$ $\Pr\left(Y(1,d)=Y(0,d)=Y(d)|X=x\right)=1$.
	\end{assumption}
	
	\begin{assumption}[Conditional monotonicity] \label{ass:monotonicity_cov}
		$\Pr\left(D(1)\ge D(0)|X=x\right)=1$ for all $x\in\mathcal{X}$.
	\end{assumption}
	
	\begin{assumption}[Conditional first stage]\label{ass:first_stage_cov}
		$\pi(1,x)-\pi(0,x)=E[D|Z=1,X=x]-E[D|Z=0,X=x]> 0$ for all $x\in\mathcal{X}$.
	\end{assumption}
	
	\begin{assumption}[Common support across $Z$] \label{ass:comsup_cov}
		$0<p(x)=\Pr(Z=1|X=x)<1$ for all $x\in\mathcal{X}$.
	\end{assumption}
	
	Assumption \ref{ass:independence_cov} consists of two conditions. The first requires the instrument to be as good as randomly assigned conditional on $X$, and thus unrelated to any factor affecting the treatment or the outcome. It implies that not only the potential outcomes and potential treatment states, but also the compliance types (which are defined by the joint potential treatment states), are independent of the instrument given $X$. It also implies that $\bar V$ is uniform on $[0,1]$ and independent of $(Z,X)$. The second condition is an exclusion restriction: $Z$ must not affect $Y$ other than through $D$, conditional on $X$. Assumption \ref{ass:monotonicity_cov} rules out the existence of defiers at every value of $X$. \citet{Vytlacil_2002} shows that this no-defiers condition is equivalent to the threshold-crossing representation \eqref{treatmodel}. Assumption \ref{ass:first_stage_cov} requires a non-zero first stage at every value of $X$ in its support. Combined with Assumptions \ref{ass:independence_cov} and \ref{ass:monotonicity_cov}, it implies $\Pr(D(1)>D(0)|X=x)>0$ for all $x$, which permits identification of the conditional LATE almost everywhere. If only the unconditional LATE were of interest, the weaker requirement $\Pr(D(1)>D(0))>0$ would suffice. Assumption \ref{ass:comsup_cov} is a common support restriction requiring that no value of $X$ perfectly predicts instrument assignment; otherwise no units comparable in terms of $X$ exist across the two instrument states at some covariate values and identification breaks down.
	
	Reconsidering the CLATE in \eqref{CLATE}, we note that
	\begin{eqnarray}\label{intMTE}
		\Delta(x,c)&=&E[Y(1)-Y(0)|X=x,\ \psi(0,x)< V \leq \psi(1,x)]\\
		&=&\frac{\int^{\psi(1,x)}_{\psi(0,x)} E\left[Y(1)-Y(0)\big|V=v,X=x\right] f_{V|X=x}(v)\,dv}{F_{V|Z=1,X=x}(\psi(1,x))-F_{V|Z=0,X=x}(\psi(0,x))}\notag,
	\end{eqnarray}
	where $E[Y(1)-Y(0)|V=v,X=x]$ is the marginal treatment effect (MTE) of \citet{Heckman_Vytlacil_1999,Heckman_Vytlacil_2005}, i.e.\ the average effect among units that are indifferent between taking and not taking the treatment at $\psi(z,x)=v$.
	
	Define the reduced form outcome regression as $\mu(Z,X)=E[Y|Z,X]$. Under Assumptions \ref{ass:independence_cov} to \ref{ass:comsup_cov}, the normalizing constant in \eqref{intMTE} satisfies $F_{V|Z=1,X=x}(\psi(1,x))-F_{V|Z=0,X=x}(\psi(0,x))=\Pr(T=c|X=x)=\pi(1,x)-\pi(0,x)$, while the integral equals $\Delta(x,c)\cdot \Pr(T=c|X=x)=\mu(1,x)-\mu(0,x)$. Therefore,
	\begin{eqnarray}\label{condLATE}
		\Delta(x,c)&=&\frac{\mu(1,x)-\mu(0,x)}{\pi(1,x)-\pi(0,x)}.
	\end{eqnarray}
	Equivalently, the CLATE may be written in terms of the efficient score functions of \citet{Tan_2006}; see \citet{Ogburn_Rotnitzky_Robins_2015} for the version conditional on covariates. Define
	\begin{eqnarray}\label{scores}
		\psi^{Y}&=&\mu(1,X)-\mu(0,X)+\frac{Z\left(Y-\mu(1,X)\right)}{p(X)}-\frac{(1-Z)\left(Y-\mu(0,X)\right)}{1-p(X)},\notag\\[2pt]
		\psi^{D}&=&\pi(1,X)-\pi(0,X)+\frac{Z\left(D-\pi(1,X)\right)}{p(X)}-\frac{(1-Z)\left(D-\pi(0,X)\right)}{1-p(X)},
	\end{eqnarray}
	so that the CLATE satisfies
	\begin{eqnarray}\label{condLATE_fr}
		\Delta(x,c)&=&\frac{E\left[\psi^{Y}\big|X=x\right]}{E\left[\psi^{D}\big|X=x\right]}.
	\end{eqnarray}
	By iterated expectations, $E[Z(Y-\mu(1,X))/p(X)|X=x]=0$ and $E[(1-Z)(Y-\mu(0,X))/(1-p(X))|X=x]=0$ under Assumption \ref{ass:comsup_cov}, and analogously for the denominator, so that $E[\psi^{Y}|X=x]=\mu(1,x)-\mu(0,x)$ and $E[\psi^{D}|X=x]=\pi(1,x)-\pi(0,x)$. The value of \eqref{scores} over \eqref{condLATE} lies in the Neyman orthogonality of the two scores with respect to the nuisance parameters $(\mu,\pi,p)$, which we exploit in Sections~\ref{sec:estimation} and \ref{sec:asymptotics}.
	
	Identification of the ICLATEs requires one further condition, which restricts the evaluation point rather than the data generating process. Write
	\begin{eqnarray}\label{Rdef}
		R=r_0(X)=\left(\tilde{X}^\top,\pi(0,X),\pi(1,X)\right)^\top, \qquad r=\left(\tilde{x}^\top,\pi',\pi''\right)^\top,
	\end{eqnarray}
	and let $d_R=\dim(\tilde X)+2$ denote the dimension of $R$ and $\mathcal{S}_R$ its support.
	
	\begin{assumption}[Support of the conditioning set]\label{ass:support}
		The evaluation point $r$ lies in the interior of $\mathcal{S}_R$ and the density $f_R$ of $R$ is bounded away from zero in a neighborhood of $r$.
	\end{assumption}
	
	Assumption \ref{ass:support} requires that units with $\tilde{X}=\tilde{x}$ and propensity scores $(\pi',\pi'')$ exist in the population, and, since $\text{ICLATE}_1$ is of interest as a function of $(\tilde x, \pi',\pi'')$, that the covariates excluded from $\tilde{X}$ generate variation in $\pi(0,\cdot)$ and $\pi(1,\cdot)$ conditional on $\tilde{X}$. If $\tilde{X}=X$, the propensity scores are deterministic functions of the conditioning variables, the conditioning on $(\pi',\pi'')$ is vacuous, and $\text{ICLATE}_1$ collapses to the CLATE. The informativeness of the ICLATE approach thus rests on $X\setminus\tilde{X}$ being non-degenerate and relevant for treatment selection, which is the same kind of variation that is required for the tests in Section~\ref{sec:extrapolation}. Whether it is present is directly verifiable from the estimated propensity scores. Note that Assumption \ref{ass:first_stage_cov} implies $\pi''>\pi'$ for any $r\in\mathcal{S}_R$, so that the complier share at the evaluation point is strictly positive.
	
	By the law of iterated expectations, the ICLATEs can be expressed as functions of the CLATE. Consider $\text{ICLATE}_1$ as defined in \eqref{firsteff} and note that
	\begin{eqnarray}\label{firsteff2}
		&&\Delta(\tilde{x},c,\pi',\pi'')=E\left[Y(1)-Y(0)|\tilde{X}=\tilde{x},T=c,Q=q\right]\notag\\
		&=& E\left[E\left[Y(1)-Y(0)|X,T=c,\pi(0,X), \pi(1,X) \right]\Big|\tilde{X}=\tilde{x},T=c,Q=q \right]\notag\\
		&=&E\left[E\left[Y(1)-Y(0)|X,T=c\right]\Big|\tilde{X}=\tilde{x}, T=c,Q=q \right]\notag\\
		&=&E\left[\Delta(X,c)\Big|\tilde{X}=\tilde{x}, T=c,Q=q \right].
	\end{eqnarray}
	The first equality follows from the law of iterated expectations because $\tilde{X}\subseteq X$; the second from the fact that $\pi(0,X)$ and $\pi(1,X)$ are deterministic functions of $X$ and may therefore be omitted from the conditioning set of the inner expectation once $X$ is conditioned on; and the third from the definition of the CLATE. In the same way,
	\begin{eqnarray}\label{secondeff2}
		\Delta(c,\pi',\pi'')&=&E\left[\Delta(X,c) \big| T=c,Q=q \right],\\
		\label{thirdeff2}
		\Delta(\tilde{x}, c)&=& E\left[\Delta(X,c) \big| T=c, \tilde{X}=\tilde{x}\right].
	\end{eqnarray}
	Theorem \ref{thm:ident} demonstrates that the ICLATEs provided in expressions \eqref{firsteff2} to \eqref{thirdeff2} are identified in the data under our IV assumptions.
	
	\begin{theorem}[Identification]\label{thm:ident}
		Let Assumptions \ref{ass:independence_cov} to \ref{ass:support} hold. Then
		\begin{eqnarray}\label{firsteff3}
			\text{ICLATE}_1&=&\frac{E\left[\mu(1,X)-\mu(0,X)\big| \tilde{X}=\tilde{x}, Q=q\right]}{E\left[\pi(1,X)-\pi(0,X)\big| \tilde{X}=\tilde{x}, Q=q\right]},
		\end{eqnarray}
		and the analogous expressions obtain for $\text{ICLATE}_2$ and $\text{ICLATE}_3$ when conditioning on $Q=q$ or on $\tilde{X}=\tilde{x}$ alone.
	\end{theorem}
	
	The proofs of this section and of Sections~\ref{sec:relation} and \ref{sec:extrapolation} are collected in Appendix~\ref{appI}.
	
	Equation \eqref{firsteff3} shows that $\text{ICLATE}_1$ is the ratio of a conditional intention-to-treat effect of the instrument on the outcome to a conditional first-stage effect of the instrument on the treatment, where both are conditioned on the coarsened information set $R$ rather than on $X$. Starting instead from the score representation \eqref{condLATE_fr} and applying the law of iterated expectations yields the equivalent doubly robust form
	\begin{eqnarray}\label{firsteff4}
		\text{ICLATE}_1=\frac{E\left[\psi^{Y}\big| R=r\right]}{E\left[\psi^{D}\big| R=r\right]},
	\end{eqnarray}
	and, for the remaining two parameters,
	\begin{eqnarray}\label{secondthird}
		\text{ICLATE}_2=\frac{E\left[\psi^{Y}\big| Q=q\right]}{E\left[\psi^{D}\big| Q=q\right]},
		\qquad
		\text{ICLATE}_3=\frac{E\left[\psi^{Y}\big| \tilde{X}=\tilde{x}\right]}{E\left[\psi^{D}\big| \tilde{X}=\tilde{x}\right]}.
	\end{eqnarray}
	
	Two features of \eqref{firsteff3} deserve comment. First, because the conditioning set fixes the values of both propensity scores, the denominator in \eqref{firsteff3} is degenerate:
	\begin{eqnarray*}
		E\left[\pi(1,X)-\pi(0,X)\big| \tilde{X}=\tilde{x}, Q=q\right]=\pi''-\pi',
	\end{eqnarray*}
	and the same holds for $\text{ICLATE}_2$. At the population level one may therefore write $\text{ICLATE}_1=E[\mu(1,X)-\mu(0,X)|R=r]/(\pi''-\pi')$, with a denominator that is known once the evaluation point is chosen. This does not hold for $\text{ICLATE}_3$, whose denominator $E[\pi(1,X)-\pi(0,X)|\tilde{X}=\tilde{x}]$ is a genuine unknown. We nevertheless retain the ratio form for all three parameters in estimation. The reason is that the empirical counterpart of $R$ is constructed from estimated propensity scores and the second step smooths over a neighborhood of $r$ of width proportional to the bandwidth, so that the estimated denominator is not exactly $\pi''-\pi'$ in finite samples. Retaining it lets the first-stage estimation error and the local smoothing bias in the numerator and the denominator offset each other, at the cost of an additional variance contribution. We return to this trade-off in Section~\ref{sec:estimation}. Second, setting $\tilde{X}=X$ returns the CLATE, so the CLATE is a boundary case of $\text{ICLATE}_1$. More generally, $\text{ICLATE}_1$ aggregates CLATEs over the covariates excluded from $\tilde{X}$, but only over those excluded covariates that leave both propensity scores unchanged. 
	
	\section{Relation to the marginal treatment effect}\label{sec:relation}
	
	The ICLATEs admit a transparent representation in terms of marginal treatment effects (MTE), which clarifies both what they hold fixed and how they differ from each other. Define
	\begin{eqnarray}\label{MTEdef}
		\Delta(x,v)=E\left[Y(1)-Y(0)\big|X=x, \bar{V}=v\right], \qquad (x,v)\in\mathcal{X}\times(0,1),
	\end{eqnarray}
	the MTE of \citet{Heckman_Vytlacil_1999,Heckman_Vytlacil_2005} expressed as a function of the normalized unobservable and the covariates.
	
	\begin{proposition}[MTE representation]\label{thm:mte}
		Let Assumptions \ref{ass:independence_cov} to \ref{ass:support} hold. Then
		\begin{eqnarray}
			\text{ICLATE}_1&=&\frac{1}{\pi''-\pi'}\int_{\pi'}^{\pi''} E\left[\Delta(X,v)\big|R=r\right]dv, \label{mte1}\\
			\text{ICLATE}_2&=&\frac{1}{\pi''-\pi'}\int_{\pi'}^{\pi''} E\left[\Delta(X,v)\big|Q=q\right]dv, \label{mte2}\\
			\text{ICLATE}_3&=&\frac{E\left[\int_{\pi(0,X)}^{\pi(1,X)}\Delta(X,v)\,dv\Big|\tilde{X}=\tilde{x}\right]}{E\left[\pi(1,X)-\pi(0,X)\big|\tilde{X}=\tilde{x}\right]}. \label{mte3}
		\end{eqnarray}
	\end{proposition}
	
	Proposition~\ref{thm:mte} makes the difference between the three LATE parameters and their relation to the MTE explicit. $\text{ICLATE}_1$ and $\text{ICLATE}_2$ average the MTE over a \emph{fixed} interval $(\pi',\pi'']$ of the unobservable, differing only in whether the averaging over covariates is restricted to $\tilde{X}=\tilde{x}$. They are therefore covariate-specific versions of the $\text{LATE}(u',u'')$ parameter of \citet{Heckman_Vytlacil_2005}. $\text{ICLATE}_3$, in contrast, averages over a \emph{covariate-dependent} interval $(\pi(0,X),\pi(1,X)]$ and weights covariate cells by the width of that interval, i.e.\ by the local complier share. Comparing $\text{ICLATE}_3$ across values of $\tilde{x}$ thus confounds effect heterogeneity in $\tilde{X}$ with shifts in both the location and the length of the range of unobservables over which the effect is averaged. The same holds for the CLATE, which corresponds to \eqref{mte3} with $\tilde{X}=X$.
	
	\section{Extrapolation to average treatment effects and testable implications}\label{sec:extrapolation}
	
	This section discusses assumptions for complier effects to be informative about conditional or unconditional average treatment effects (ATE) in the total population, and what the data can say about those assumptions. 
   To this end, let $\Delta(x)=E[Y(1)-Y(0)|X=x]$, $\Delta(\tilde{x})=E[Y(1)-Y(0)|\tilde{X}=\tilde{x}]$, and $\Delta=E[Y(1)-Y(0)]$ denote the conditional and unconditional ATEs. We first consider an assumption that rules out effect heterogeneity in unobservables.
	
	\begin{assumption}[No effect heterogeneity in the unobservable $V$]\label{ass:homoV}
		$\Delta(x,v)=\Delta(x)$ for all $v\in(0,1)$ and all $x\in\mathcal{X}$.
	\end{assumption}
	
	Assumption \ref{ass:homoV} restricts only the dependence of the effect on $V$ and leaves heterogeneity across $X$ unrestricted. It is closely related to an assumption considered in \citet{Angrist_FernandezVal_2013} and \citet{Aronow_Carnegie_2013}, namely equality of average effects across compliance types given $X$. Under this assumption, $\Delta(x,c)=\Delta(x)$ by \eqref{intMTE}, so that the CLATE extrapolates pointwise and, by \eqref{condLATE},
	\begin{eqnarray}\label{cate_ident}
		\Delta(\tilde{x})=E\left[\frac{\mu(1,X)-\mu(0,X)}{\pi(1,X)-\pi(0,X)}\Big|\tilde{X}=\tilde{x}\right], \quad
		\Delta=E\left[\frac{\mu(1,X)-\mu(0,X)}{\pi(1,X)-\pi(0,X)}\right],
	\end{eqnarray}
	see the discussion in \citet{Angrist_FernandezVal_2013} and \citet{Aronow_Carnegie_2013}. 
	
	Relating this restriction on the CLATE given $X$ to our ICLATE framework, it is important to note that Assumption \ref{ass:homoV} does not rule out the dependence of $\text{ICLATE}_1$ on $(\pi',\pi'')$ and thus, unobservables. By \eqref{mte1} it gives $\text{ICLATE}_1=E[\Delta(X)|R=r]$, which still varies with $(\pi',\pi'')$ at fixed $\tilde{x}$, because conditioning on the propensity scores restricts $X\setminus\tilde{X}$ to a level set that moves with them such that not all elements in $X$ are controlled for. For this reason, we subsequently consider a complementary restriction that rules out effect heterogeneity across excluded covariates.
	
	\begin{assumption}[Effect homogeneity across excluded covariates]\label{ass:homoX}
		$\Delta(x,v)=\Delta(\tilde{x},v)$ for all $v\in(0,1)$ and all $x\in\mathcal{X}$ with $\tilde{x}\subseteq x$, where $\Delta(\tilde{x},v)=E\left[Y(1)-Y(0)\big|\tilde{X}=\tilde{x},\bar{V}=v\right]$.
	\end{assumption}
	
	Assumption \ref{ass:homoX} is the mirror image of Assumption \ref{ass:homoV} in that it permits arbitrary effect heterogeneity in $V$ and imposes homogeneity in the covariates excluded from $\tilde{X}$. Neither restriction implies the other, but their joint satisfaction is testable in the data, as formally stated in the proposition below.
	
	\begin{proposition}[Testable implication]\label{prop:test}
		Let Assumptions \ref{ass:independence_cov} to \ref{ass:support} hold.  Suppose $\Delta(x,v)=\Delta(\tilde{x})$ for all $v\in(0,1)$ and all $x\in\mathcal{X}$ with $\tilde{x}\subseteq x$, that is, suppose Assumptions \ref{ass:homoV} and \ref{ass:homoX} both hold. Then $\text{ICLATE}_1=\text{ICLATE}_3=\Delta(\tilde{x})$ for every $(\pi',\pi'')\in\mathcal{S}_R(\tilde{x})=\{(\pi',\pi''):(\tilde{x}^\top,\pi',\pi'')^\top\in\mathcal{S}_R\}$, so that
		\begin{eqnarray}\label{H0test}
			H_0: \quad \Delta(\tilde{x},c,\pi',\pi'')=\Delta(\tilde{x},c,\dot{\pi}',\dot{\pi}'') \quad \text{for all } (\pi',\pi''),(\dot{\pi}',\dot{\pi}'')\in\mathcal{S}_R(\tilde{x})
		\end{eqnarray}
		holds. Neither assumption  on its own implies $H_0$.
	\end{proposition}
	
	Each $\Delta(\tilde{x},c,\pi',\pi'')$ is identified by Theorem~\ref{thm:ident}, so \eqref{H0test} is a restriction on the observable data distribution and can be assessed by comparing estimates across a grid of propensity score values at a fixed $\tilde{x}$. A rejection rejects the conjunction of homogeneity in the unobservable and homogeneity across the excluded covariates, and does not by itself attribute the failure to either. Maintaining Assumption \ref{ass:homoX} a priori turns \eqref{H0test} into a test of selection on unobservables; maintaining Assumption \ref{ass:homoV} turns it into a test of effect heterogeneity across the excluded covariates.

	When imposing Assumption \ref{ass:homoX}, the MTE no longer varies with the covariates excluded from $\tilde{X}$, so the averaging over them in \eqref{mte1} becomes irrelevant and $\text{ICLATE}_1$ is an average of the single function $\Delta(\tilde{x},\cdot)$ over $(\pi',\pi'']$. Theorem~\ref{thm:mtecurve} states this formally and shows that $\Delta(\tilde{x},\cdot)$ is then identified over the range of the unobservable that the propensity scores trace out. It also pins down what \eqref{H0test} tests: if the average of $\Delta(\tilde{x},\cdot)$ is the same over every interval, the function is constant in $v$, so that the effect does not vary with the unobservable. 	
	
	\begin{theorem}[MTE curve at the level of $\tilde{X}$]\label{thm:mtecurve}
		Let Assumptions \ref{ass:independence_cov} to \ref{ass:support} and \ref{ass:homoX} hold. Define $M(\tilde{x},v)=\int_0^v \Delta(\tilde{x},u)\,du$. Then, for every $(\pi',\pi'')\in\mathcal{S}_R(\tilde{x})$,
		\begin{eqnarray}\label{diffquot}
			\text{ICLATE}_1=\Delta(\tilde{x},c,\pi',\pi'')=\frac{M(\tilde{x},\pi'')-M(\tilde{x},\pi')}{\pi''-\pi'}.
		\end{eqnarray}
		Consequently, if $\Delta(\tilde{x},\cdot)$ is continuous and if for some $\pi'_0$ the set $\{\pi'': (\pi'_0,\pi'')\in\mathcal{S}_R(\tilde{x})\}$ contains an interval $\mathcal{V}(\tilde{x})$, then $\Delta(\tilde{x},v)$ is identified for all $v$ in the interior of $\mathcal{V}(\tilde{x})$.
	\end{theorem}
	
	Proposition~\ref{prop:additivity} presents a further implication of Theorem~\ref{thm:mtecurve}, which we refer to as interval additivity: the effect over the union of two adjacent intervals of the propensity scores is the length-weighted average of the effects over the two pieces. This is a testable restriction on Assumption \ref{ass:homoX} itself. Without that assumption it need not hold. The three intervals of the unobservable always differ, but conditioning on $R=r$ also restricts the excluded covariates to a level set that moves with $(\pi',\pi'')$, so that the three terms in \eqref{additivity} below average the marginal treatment effect over three different distributions of $X\setminus\tilde{X}$ and the weighted average no longer recovers the effect over the union. Testing it requires three pairs with chained endpoints at the same $\tilde{x}$, which is a stronger support requirement than Assumption \ref{ass:support} and should be verified against the estimated propensity scores.
	
	\begin{proposition}[Interval additivity]\label{prop:additivity}
		Let the assumptions of Theorem~\ref{thm:mtecurve} hold and let $\pi'<\pi''<\pi'''$ be such that $(\pi',\pi'')$, $(\pi'',\pi''')$ and $(\pi',\pi''')$ all belong to $\mathcal{S}_R(\tilde{x})$. Then
		\begin{eqnarray}\label{additivity}
			(\pi'''-\pi')\,\Delta(\tilde{x},c,\pi',\pi''')=(\pi''-\pi')\,\Delta(\tilde{x},c,\pi',\pi'')+(\pi'''-\pi'')\,\Delta(\tilde{x},c,\pi'',\pi''').
		\end{eqnarray}
	\end{proposition}
	
	Theorem~\ref{thm:mtecurve} and Propositions~\ref{prop:test} and \ref{prop:additivity} have an $\text{ICLATE}_2$ counterpart under the stronger restriction that the effect does not vary with any covariate, $\Delta(x,v)=\Delta(v)$ for all $x\in\mathcal{X}$. Writing $M(v)=\int_0^v\Delta(u)\,du$ one obtains $\text{ICLATE}_2=\left\{M(\pi'')-M(\pi')\right\}/(\pi''-\pi')$, so that the population marginal treatment effect curve is identified over the range the propensity scores trace out, constancy of $\text{ICLATE}_2$ across those scores is equivalent to no selection on unobservables, and the weighted average of Proposition~\ref{prop:additivity} again recovers the effect over the union. The restriction is stronger, but the practical advantage is that $\text{ICLATE}_2$ conditions on two variables rather than on $\dim(\tilde X)+2$ and is therefore estimated at a faster rate. 
Section~\ref{sec:asymptotics} provides joint inference
at a fixed finite set of prespecified evaluation points.
The Wald statistic in \eqref{cf:finitewald} therefore
permits tests of constancy and interval additivity
on the selected grid, provided the corresponding
support and inference conditions hold at every point.

	\section{Estimation}\label{sec:estimation}
	
	We estimate $\text{ICLATE}_1$ based on the doubly robust representation \eqref{firsteff4}. Write $A_i=\psi^{Y}(W_i)$ and $B_i=\psi^{D}(W_i)$ for the two scores of \eqref{scores} evaluated at observation $i$, that is,
	\begin{align}\label{Ai}
		A_i&=\mu(1,X_i)-\mu(0,X_i)+\frac{Z_i\left(Y_i-\mu(1,X_i)\right)}{p(X_i)}-\frac{(1-Z_i)\left(Y_i-\mu(0,X_i)\right)}{1- p(X_i)},
		\\
		\label{Bi}
		B_i &= \pi(1,X_i)-\pi(0,X_i)+\frac{Z_i\left(D_i-\pi(1,X_i)\right)}{ p(X_i)}-\frac{(1-Z_i)\left(D_i-\pi(0,X_i)\right)}{1- p(X_i)},
	\end{align}
	and recall $R_i=(\tilde X_i^\top, \pi(0,X_i), \pi(1, X_i))^\top$ and $r=(\tilde x^\top, \pi', \pi'')^\top$ from \eqref{Rdef}. Writing $\gamma_A(r)=E[A_i|R_i=r]$ and $\gamma_B(r)=E[B_i|R_i=r]$, Theorem~\ref{thm:ident} gives $\text{ICLATE}_1=\gamma_A(r)/\gamma_B(r)$. Estimation proceeds in two steps: the nuisance functions $(\mu,\pi,p)$ are estimated by machine learning methods with cross-fitting, and the resulting scores are regressed on the estimated $R_i$ by local linear regression. This structure follows \citet{Abrevaya_Hsu_Lieli_2015}, \citet{Zimmert_Lechner_2019}, and \citet{Fan_Hsu_Lieli_Zhang_2019}, with the additional complication that here the conditioning variables themselves contain estimated functions. 
    The procedure for $\text{ICLATE}_1$ is as follows.
	\begin{changemargin}{0.6cm}{0cm}
		\begin{itemize}
			\item[Step 1.]  {\bf(Nuisance functions)}
			For a fixed integer $L\geq2$ and $n\geq L$, randomly partition the observation indices independently of the data into $L$ groups $\mathcal{I}_l$, $l=1,\dots,L$, of approximately equal size. For each $l$, the estimators $\hat\mu_l(Z_i,X_i)$, $\hat p_l(X_i)$, and $\hat\pi_l(Z_i,X_i)$ of $\mu(Z_i,X_i)$, $p(X_i)$, and $\pi(Z_i,X_i)$ use only observations outside $\mathcal{I}_l$ and satisfy Assumption~\ref{A1stCLATE} below.
			The treatment propensity estimators additionally satisfy Assumption~\ref{Aindex}, because they enter the conditioning index.
			 Write $\hat\mu_{-i}=\hat\mu_l$, $\hat p_{-i}=\hat p_l$, and $\hat\pi_{-i}=\hat\pi_l$ for $i\in\mathcal{I}_l$. Estimate $A_i$ and $B_i$ by
			\begin{align*}
				\hat A_{i}&=\hat\mu_{-i}(1,X_i)-\hat\mu_{-i}(0,X_i)  +\frac{Z_i\left(Y_i-\hat\mu_{-i}(1,X_i)\right)}{\hat p_{-i}(X_i)}-\frac{(1-Z_i)\left(Y_i-\hat\mu_{-i}(0,X_i)\right)}{1-\hat p_{-i}(X_i)},\\[2pt]
				\hat B_i &=\hat \pi_{-i}(1,X_i)-\hat\pi_{-i}(0,X_i) +\frac{Z_i\left(D_i-\hat\pi_{-i}(1,X_i)\right)}{\hat p_{-i}(X_i)}-\frac{(1-Z_i)\left(D_i-\hat\pi_{-i}(0,X_i)\right)}{1-\hat p_{-i}(X_i)},
			\end{align*}
			for $i=1,\dots,n$, and set $\hat r(X_i)=(\tilde X_i^\top, \hat\pi_{-i}(0,X_i), \hat\pi_{-i}(1,X_i))^\top$.
			
			\item[Step 2.] {\bf (Local linear regression)}
			For $\Psi\in\{A,B\}$, regress $\hat \Psi_{i}$ on $\hat r(X_i)$ locally at $r$:
			\begin{align*}
				\hat\gamma_\Psi(r) =&\ \hat{E}\big[\hat\Psi_{i} \big| \hat r(X_i) = r\big] = \hat a,\quad \mbox{where}
\\
(\hat a, \hat b_0, \hat b_1, \hat b_2) =&\ \arg\min_{a, b_0, b_1, b_2}
				\sum_{i=1}^n \mathcal{K}_h\left(\hat r(X_i) - r\right) \Big(
				\hat \Psi_{i} - a- b_0^\top(\tilde X_i - \tilde x)\\
				&\qquad - b_1\big(\hat\pi_{-i}(0, X_i) - \pi'\big) - b_2\big(\hat\pi_{-i}(1, X_i) - \pi''\big)
				\Big)^2,
			\end{align*}
			where $\mathcal K_h(u)=h^{-d_R}\prod_{j=1}^{d_R}k(u_j/h)$
			denotes the product kernel built from a univariate kernel $k$ and a bandwidth $h$, as specified in Assumption~\ref{Akernel} below. In the implementation we use coordinate-specific bandwidths $h_j=c_jh$, with the scales $c_j$ estimated from the data; the bandwidth discussion in Section~\ref{sec:asymptotics} shows that the results are unaffected.
		\end{itemize}
	\end{changemargin}
	
	The estimator is $\widehat{\text{ICLATE}}_1 = \hat\gamma_A(r)/\hat\gamma_B(r)$. Estimators of $\text{ICLATE}_2$ and $\text{ICLATE}_3$ are obtained by replacing $\hat r(X_i)$ with $(\hat\pi_{-i}(0,X_i),\hat\pi_{-i}(1,X_i))^\top$ and with $\tilde X_i$, respectively. 
	Theorem~\ref{TAsym23} establishes the corresponding asymptotic results. For $\text{ICLATE}_3$, the conditioning variables are observed, so no generated-index accuracy or continuity condition is required; see also \citet{Zimmert_Lechner_2019} and \citet{Fan_Hsu_Lieli_Zhang_2019}.
	
	Since $\gamma_B(r)=\pi''-\pi'$ is known at the evaluation
point, an alternative is
$\widetilde{\text{ICLATE}}_1=\hat\gamma_A(r)/(\pi''-\pi')$.
Under the conditions of Section~\ref{sec:asymptotics},
its asymptotic variance under $\sqrt{nh^{d_R}}$
normalization is
${R_k^{d_R}\operatorname{var}(A\mid R=r)}
 /({f_R(r)(\pi''-\pi')^2})$.
For the same kernel and bandwidth, our ratio estimator
has smaller asymptotic variance if and only if
$ 2\,\text{ICLATE}_1(r)\operatorname{cov}(A,B\mid R=r)
 >
 \text{ICLATE}_1(r)^2\operatorname{var}(B\mid R=r)$.
Neither estimator therefore uniformly dominates the other.
We retain the ratio form for all three ICLATEs, using
common local linear weights for the numerator and denominator.
	
	\section{Asymptotic theory}\label{sec:asymptotics}

We establish pointwise inference for the pooled local linear estimator in Section~\ref{sec:estimation}. The proof distinguishes the estimation error of the nuisance functions  in the orthogonal scores from error in the conditioning index. 
Cross-fitting allows us to analyze each fold's contribution to the local linear estimator conditional on the corresponding training sample.
A local continuity condition controls the conditional projection of the index bias and allows that bias to be nonzero. Under the stated accuracy conditions, the estimator has the same first-order expansion as the estimator using the true scores and index.

\subsection{Sampling, kernels, and local regularity}

Let $W_i=(Y_i,D_i,Z_i,X_i)$, $d=d_R$, and $R_i=r_0(X_i)$.
The observations are independent and identically distributed.
Let 
\[
 \theta(r)=\text{ICLATE}_1(r)=\frac{\gamma_A(r)}{\gamma_B(r)},
 \qquad \gamma_\Psi(t)=E[\Psi\mid R=t],\quad \Psi\in\{A,B\}.
\]
The scores $A$ and $B$ are defined in \eqref{Ai}--\eqref{Bi}. Throughout this section, $r$ is a fixed interior evaluation point.  The number of folds $L$ is fixed, the partition is independent of the observations, and $n_l/n\to\omega_l\in(0,1)$, where $n_l=|\mathcal I_l|$. 
Let $\mathcal F_{-l}$ contain the observations outside $\mathcal I_l$. 
All preprocessing and tuning for fold $l$ use only that training sample. For a fresh observation $W$, independent of the data, write $E_l[\cdot]=E[\cdot\mid\mathcal F_{-l}]$. In particular, fitted functions are held fixed inside $E_l$.

\begin{assumption}[Kernel and bandwidth]\label{Akernel}
The product kernel is $\mathcal K_h(t)=h^{-d}K(t/h)$, where $K(u)=\prod_{j=1}^d k(u_j)$. Either $k$ is a continuously differentiable symmetric probability density supported on $[-1,1]$, with $\int u^2k(u)\,du>0$, or $k$ is the standard Gaussian density. The deterministic bandwidth satisfies $h=c_h n^{-\eta}$, where $c_h>0$ and
$ {(d+4)^{-1}}<\eta<{d^{-1}}$.
Thus $nh^d\to\infty$ and $nh^{d+4}\to0$.
\end{assumption}

Let $\tau_n=\sqrt{nh^d}$, $R_k=\int k(u)^2\,du$, and $\mu_2(k)=\int u^2k(u)\,du$. The bandwidth restriction makes the local linear bias $O(h^2)$ negligible relative to $\tau_n^{-1}$. 
\begin{assumption}[Local regularity and moments]\label{Areg}
There is a bounded open neighborhood $\mathcal N$ of $r$, with closure contained in the interior of $\mathcal S_R$, such that:
\begin{enumerate}[label=(\roman*)]
 \item The density of $R$, $f_R$, is continuous, bounded, and bounded away from zero on ${\mathcal N}$.
 \item $\gamma_A$ and $\gamma_B$ have continuous second derivatives on ${\mathcal N}$.
 \item The conditional covariance matrix of $(A,B)^\top$ is continuous at $r$, and
$ \sigma_{AB}^2(r)=\operatorname{var}(A-\theta(r)B\mid R=r)>0.$
 \item For some $\nu>0$,
 $ \operatorname*{ess\,sup}_{t\in\mathcal N}
 E[|A|^{2+\nu}+|B|^{2+\nu}\mid R=t]<\infty.$
 Also $E[|A|+|B|]<\infty$; for a Gaussian kernel, assume in addition that $E[|A|^{2+\nu}+|B|^{2+\nu}]<\infty$.
 
 \item Let
$\sigma_z^2(x)=\operatorname{var}(Y\mid Z=z,X=x)$.
For a fixed constant $C_\sigma<\infty$, $ \max_{z\in\{0,1\}}\sigma_z^2(x)\leq C_\sigma$ for $P_X$-almost every $x$ satisfying
$r_0(x)\in\mathcal N$.
For the Gaussian kernel, impose this bound for
$P_X$-almost every $x\in\mathcal X$.

\end{enumerate}
\end{assumption}

Part (v) allows the conditional outcome variance to vary with the
instrument and covariates. Its upper bound lets us control the
contribution of the instrument propensity score error using the
partial $L_2$ norm defined below.

\subsection{Partial $L_2$ rates for nuisance estimation}

Since our nonparametric estimators are generally non-regular, i.e., the convergence rate is slower than root-$n$, we state the nuisance-rate conditions using a density-weighted partial $L_2$ norm, analogous to the norm in \citet{CL}. It holds the true conditioning index fixed and integrates over the remaining variation in the observation. We require the partial $L_2$ rate consistency and product-rate conditions, which are standard in the DML literature \citep{Chernozhukov_Chetverikov_Demirer_Duflo_Hansen_Newey_2017, CL}.

For $t\in\mathcal N$, let $P_{W\mid R=t}$ be a regular conditional distribution.
For a scalar function $g$ of $W$, define
\begin{equation}\label{cf:partialnorm}
 \|g\|_{l,t,2}^2 =
 \int g(w)^2\,P_{W|R=t}(dw)f_R(t)
 =f_R(t)E_l[g(W)^2\mid R=t].
\end{equation}
The fitted function is held fixed conditional on $\mathcal F_{-l}$. 
These objects are defined for almost every $t$. 
Since $R=r_0(X)$ is a deterministic function of $X$,
the joint distribution of $(R,X)$ has no density with
respect to full-dimensional Lebesgue measure.
Definition~\eqref{cf:partialnorm} instead uses the
conditional distribution given $R=t$ and requires
a density only for $R$.
If $W$ admits coordinates $(R,U)$ with joint density $f_{RU}$ and $g(W)=\bar g(R,U)$, then \eqref{cf:partialnorm} equals $\int\bar g(t,u)^2f_{RU}(t,u)\,du$, the usual partial-$L_2$ form. 
By Assumption~\ref{Areg}, the factor $f_R(t)$ is bounded and bounded away from zero on this neighborhood; omitting it therefore yields equivalent rate conditions. 

Use the locally uniform partial norm
 $\|g\|_{l,\mathcal N,2}
 =\operatorname*{ess\,sup}_{t\in\mathcal N}\|g\|_{l,t,2}$.
Write $e_{\mu,z,l}=\hat\mu_l(z,\cdot)-\mu(z,\cdot)$, $e_{\pi,z,l}=\hat\pi_l(z,\cdot)-\pi(z,\cdot)$, and $e_{p,l}=\hat p_l-p$, and define
\begin{align*}
 \mathcal{E}_{\mu,l}&=\sum_{z=0}^1\|e_{\mu,z,l}\|_{l,\mathcal N,2}, \qquad
 \mathcal{E}_{\pi,l}=\sum_{z=0}^1\|e_{\pi,z,l}\|_{l,\mathcal N,2},\qquad
 \mathcal{E}_{p,l}=\|e_{p,l}\|_{l,\mathcal N,2}.
\end{align*}

\begin{assumption}[Cross-fitted nuisance functions]\label{A1stCLATE}
For some fixed $\varepsilon>0$, $\varepsilon\leq p(X)\leq1-\varepsilon$ almost surely and $\hat p_l(X)\in[\varepsilon,1-\varepsilon]$, $\hat\pi_l(z,X)\in[0,1]$ almost surely, for every fold and instrument value. The fitted functions are $\mathcal F_{-l}$-measurable and
\begin{align}\label{cf:partialrates}
 \max_{l\leq L}(\mathcal{E}_{\mu,l}+\mathcal{E}_{\pi,l}+\mathcal{E}_{p,l})=o_p(1),\qquad
 \tau_n\max_{l\leq L}\mathcal{E}_{p,l}(\mathcal{E}_{\mu,l}+\mathcal{E}_{\pi,l})=o_p(1).
\end{align}
For the Gaussian kernel, assume additionally that, for some fixed $C_0\geq0$,
\begin{equation}\label{cf:partialtail}
 \max_{l\leq L}E_l\left[
 \sum_{z=0}^1\{e_{\mu,z,l}(X)^2+e_{\pi,z,l}(X)^2\}
 +e_{p,l}(X)^2\right]
 =O_p(n^{C_0}).
\end{equation}
Condition \eqref{cf:partialtail} is not required for a compactly supported kernel.
\end{assumption}

\subsection{Generated-regressor error}
For each fold $l$, let $\hat r_l(x) =(\tilde x^\top,\hat\pi_l(0,x),\hat\pi_l(1,x))^\top.$
Define the population index-bias functions
\begin{align*}
 \rho_\Psi(X)=E[\Psi\mid X]-\gamma_\Psi(R),\qquad \Psi\in\{A,B\}.
\end{align*}
By iterated expectations, $E[\rho_\Psi(X)\mid R]=0$.
For the denominator score, 
\begin{equation}\label{cf:rhoB}
 E[B\mid X]=\pi(1,X)-\pi(0,X)=\gamma_B(R),
 \qquad \rho_B(X)=0\quad\text{almost surely}.
\end{equation}

For each deterministic candidate index function $q$, define
\[
 \tau_A(t,q)=E[\rho_A(X)\mid q(X)=t] = E\left[ E[A\mid X] - E[A\mid R]\mid q(X) = t\right].
\]
Here $t$ is the evaluation point, and $q$ specifies
the conditioning index. 
The population index-bias function $\rho_A$ is defined using the true index $R=r_0(X)$ and does not depend on $q$.
As $q$ varies, only the conditioning index in $\tau_A(t,q)=E[\rho_A(X)\mid q(X)=t]$ changes.
Since $E[\rho_A(X)\mid R]=0$,
$ \tau_A(t,r_0)=0$ for $P_R\text{-almost every }t$.

\begin{assumption}[Index accuracy and local continuity]\label{Aindex}
Let  
\[
a_{l,n}=\|\hat r_l-r_0\|_\infty  =\max_{z\in\{0,1\}}\sup_{x\in\mathcal X}   |\hat\pi_l(z,x)-\pi(z,x)|.
\]
For a deterministic sequence $a_n\downarrow0$,
\begin{equation}\label{cf:indexaccuracy}
 \max_{l\leq L}a_{l,n}=O_p(a_n),\qquad
 a_n/h\to0,\qquad \tau_n a_n\to0.
\end{equation}
There are deterministic classes $\mathcal R_n$
of measurable index functions
$q(x)=(\tilde x^\top,q_0(x),q_1(x))^\top$,
with $q_z:\mathcal X\to[0,1]$ for $z=0,1$, a constant $C_\rho<\infty$ independent of $n$, and a fixed neighborhood $\mathcal N_0$ of $r$ with $\overline{\mathcal N}_0\subset\mathcal N$, such that $r_0\in\mathcal R_n$,
\begin{align}
 \Pr\{\hat r_l\in\mathcal R_n\text{ for every }l\leq L\}\to1,
\label{Emem}
\end{align}
and, for every $q\in\mathcal R_n$,
\begin{equation}\label{cf:indexstability}
 1\{q(X)\in\mathcal N_0\}
 \left|E[\rho_A(X)\mid q(X)]\right|
 \leq C_\rho\|q-r_0\|_\infty
 \qquad\text{almost surely},
\end{equation}
where
$\|q-r_0\|_\infty
=\sup_{x\in\mathcal X}\|q(x)-r_0(x)\|_\infty$.
\end{assumption}

Condition \eqref{cf:indexstability} imposes local continuity
of the projection of the index bias relative to the true index.
Together with \eqref{Emem} and the uniform accuracy bound in
\eqref{cf:indexaccuracy}, it bounds the conditional mean of
the index bias given each fitted index locally by $O_p(a_n)$,
uniformly over folds.

A sufficient, stronger formulation of \eqref{cf:indexstability} is the fixed-evaluation-point Lipschitz condition
\[
 |\tau_A(t,q_1)-\tau_A(t,q_2)|
 \leq C_\rho\|q_1-q_2\|_\infty,
 \qquad t\in\mathcal N_0,
\]
for all $q_1,q_2\in\mathcal R_n$.
Since $E[\rho_A(X)\mid R]=0$, we take $\tau_A(t,r_0)=0$ for every $t\in\mathcal N_0$.
Taking $q_2=r_0$ gives \eqref{cf:indexstability}.
This is the form used in \citet[Assumption~4]{Mammen_Rothe_Schienle_2012}, and \citet[Assumption~4]{MRS2016} give a related condition evaluated at the respective random indices. We use only the bound \eqref{cf:indexstability}, not those papers' stochastic expansions or additional derivatives of the conditional projection.
As discussed in \citet[p.~1151]{MRS2016},
such conditions can be verified through suitable
bounds on the change in the relevant densities and
a positive lower bound on the candidate-index density.
These are restrictions on the population distribution
and the candidate class, separate from the convergence
rate of the first-stage estimator.
The special case $\rho_A=0$ satisfies it automatically,
but is not required.

The class $\mathcal R_n$ specifies the candidate index
functions over which \eqref{cf:indexstability} is imposed.
It plays a role analogous to $\mathcal R_n^*$ in
\citet[Assumptions~3--4]{MRS2016}
and the smoothness classes used in
\citet[Assumption~8]{LeePMGR}.
As in those papers, the generated regressors must belong with
probability approaching one to a deterministic class
on which the continuity restriction holds.
Unlike their empirical-process arguments, our pointwise
cross-fitted proof requires no covering-number bound
on this class.
The choice of $\mathcal R_n$ may reflect the first-stage
specification, but membership and projection continuity
must be verified for the same class.

The conditions in \eqref{cf:indexaccuracy} have distinct roles. 
The requirement $a_n=o(h)$ controls centered kernel perturbations and is also familiar from the above mentioned references.
Our additional condition $\tau_n a_n\to0$ is sufficient for oracle equivalence: it makes the generated-index contributions negligible by controlling the uniform first-stage error.
The cited papers instead retain smoothed first-stage contributions and can exploit their faster convergence under additional structure.
Our pointwise cross-fitted argument avoids entropy
conditions and derivatives of fitted nuisance functions.
Thus the assumptions follow established generated-regressor
arguments, with a sufficient, rather than minimal, accuracy requirement.
For example, under the conditions of Lemma~\ref{LSUZ} in Appendix~\ref{Athems}, logistic lasso
satisfies the uniform index-accuracy requirement \eqref{cf:indexaccuracy}.
Appendix~\ref{Athems} also gives a class of logistic index functions
for which the coefficient bound in the lemma implies the membership
condition \eqref{Emem}. Projection continuity \eqref{cf:indexstability}
is a separate restriction on that class.

Appendix~\ref{app:simcontinuity} verifies
Assumptions~\ref{Areg}--\ref{Aindex} for the simulation design
in Section~\ref{sec:simulation}, under the stated regularity
conditions for the parametric nuisance estimators.
In particular, it establishes the required nuisance rates and
verifies candidate-class membership and projection continuity
for the same class of probit indices, covering both zero and
nonzero population index bias.

\begin{remark}[Nuisance estimators and admissible rates.]
The nuisance functions may be estimated by parametric,
nonparametric, or machine-learning methods satisfying
Assumption~\ref{A1stCLATE}.
The treatment propensity estimators must additionally
satisfy Assumption~\ref{Aindex}, because they enter the
conditioning index.
If $h\asymp n^{-\eta}$ and $a_n=n^{-\kappa}$, the two requirements are $\kappa>\max\{\eta,(1-d\eta)/2\}$.
For the bandwidth exponent $\eta=1.2/(d+4)$ used in Section~\ref{sec:simulation},
the combined restrictions $a_n/h\to0$ and $\tau_n a_n\to0$
require $\kappa>3/10$, $\kappa>17/70$, and $\kappa>1/5$
when $d=2,3,4$, respectively.
In particular, a uniform $O_p(n^{-1/4})$ rate suffices
for $d=3$ and $d=4$, but not for $d=2$ under this
bandwidth rule.
Uniformly root-$n$ consistent first stages satisfy
\eqref{cf:indexaccuracy} whenever $\eta<1/2$;
for example, correctly specified fixed-dimensional
logit estimators with bounded regressors and root-$n$
coefficient consistency have this uniform rate.
For high-dimensional sparse logit models,
Lemma~\ref{LSUZ} gives
$a_n=s_n\sqrt{\log(2p_n)/n}$ under its coefficient-error
and bounded-regressor conditions.
For $d=3$ and $h\asymp n^{-6/35}$,
\eqref{lassocond} reduces to $s_n\sqrt{\log(2p_n)}=o(n^{9/35})$.
Kernel, series, and other machine-learning estimators
are also admissible when their errors satisfy the
required norms and rates.
An ordinary $L_2$ prediction rate alone does not
establish the uniform index-accuracy condition,
and projection continuity must be verified separately.
\end{remark}

\subsection{Pointwise inference}

\begin{theorem}[Asymptotic normality]\label{TAsym1}
Suppose Assumptions~\ref{ass:independence_cov}--\ref{ass:support} and \ref{Akernel}--\ref{Aindex}, and the sampling and cross-fitting conditions above, hold. For the pooled estimator in Section~\ref{sec:estimation},
\begin{align}\label{cf:scoreexpansion}
 \hat\gamma_\Psi(r)-\gamma_\Psi(r)
&=\frac1n\sum_{i=1}^n\phi_{\Psi i}(r)+o_p(\tau_n^{-1}),\ \mbox{where}\\
  \phi_{\Psi i}(r)&=\frac{\mathcal K_h(R_i-r)}{f_R(r)}\big(\Psi_i-\gamma_\Psi(R_i)\big),
 \qquad \Psi\in\{A,B\}. \notag
\end{align}
Consequently, for $\hat\theta(r)=\hat\gamma_A(r)/\hat\gamma_B(r)$,
\begin{align}
 \tau_n\{\hat\theta(r)-\theta(r)\}
&=\frac{\tau_n}{n}\sum_{i=1}^n
 \frac{\phi_{Ai}(r)-\theta(r)\phi_{Bi}(r)}{\gamma_B(r)}+o_p(1) \label{asylin}\\
&\xrightarrow{d}\ N(0,V_1(r)),
 \ \ \mbox{where}\ V_1(r)=\frac{R_k^d\sigma_{AB}^2(r)}{f_R(r)\gamma_B(r)^2}. \label{asynorm}
\end{align}

\end{theorem}

For $i\in\mathcal I_l$, let $\hat R_i=\hat r_l(X_i)$. 
A convenient variance estimator is
\begin{equation}\label{varest}
 \hat V_1(r)=
 \frac{n h^{d}}{\hat\gamma_B(r)^2}
 \sum_{i=1}^n \tilde w_i(r)^2
 \{\hat A_i-\hat\theta(r)\hat B_i\}^2,
\end{equation}
where
\begin{equation}\label{llweights}
 \tilde w_i(r)=\mathcal K_h(\hat R_i-r)\,
 e_1^\top S(r)^{-1}\big(1,(\hat R_i-r)^\top\big)^\top,\quad
 S(r)=\sum_{i=1}^n\mathcal K_h(\hat R_i-r)
 \big(1,(\hat R_i-r)^\top\big)^\top\big(1,(\hat R_i-r)^\top\big),
\end{equation}
are the influence weights of the local linear fit that defines $\hat\theta(r)$. They satisfy $\sum_{i=1}^n\tilde w_i(r)=1$ and reproduce linear functions exactly.

\begin{proposition}[Feasible inference]
\label{cf:varianceprop}
Under the conditions of Theorem~\ref{TAsym1},
$\hat V_1(r)\xrightarrow{p}V_1(r)$.
Hence a pointwise asymptotic $(1-\alpha)$ confidence
interval is
 $\hat\theta(r)\ \pm\ z_{1-\alpha/2}
 \sqrt{{\hat V_1(r)}/{(nh^d)}}$,
where $z_{1-\alpha/2}$ is the $(1-\alpha/2)$ quantile of the standard normal distribution.
\end{proposition}

The variance estimator uses the cross-fitted estimated scores and consistently estimates the variance of the leading oracle term. Under the stated nuisance-rate conditions, no additional first-order adjustment for score estimation is required. 

We next consider the choice of bandwidth.
The preceding results also hold for coordinate-specific
bandwidths $h_j=c_jh$, with fixed $c_j>0$, using the
corresponding product kernel and replacing $h^d$ by
$H=\prod_{j=1}^d h_j$.
They continue to hold when
$h_j=c\sigma_jn^{-\eta}$ is replaced by
$\hat h_j=c\hat\sigma_jn^{-\eta}$, where $c>0$ is fixed,
$\sigma_j^2=\operatorname{var}(R_j)>0$, and
$\hat\sigma_j$ is the sample standard deviation of
$\{\hat R_{ij}\}_{i=1}^n$, provided $E[R_j^4]<\infty$
and the conditions of Theorem~\ref{TAsym1} hold. 
The feasible variance and confidence interval then use
$\hat H=\prod_{j=1}^d\hat h_j$ in place of $h^d$.
This covers the undersmoothed rule-of-thumb bandwidth
used in Sections~\ref{sec:simulation}
and~\ref{sec:application}.

To reduce sensitivity to the random fold assignment, we follow
\citet[Section~3.4]{Chernozhukov_Chetverikov_Demirer_Duflo_Hansen_Newey_2017} and repeat the
estimation procedure over a fixed number $S$ of random
$L$-fold partitions of the same full sample.
For each partition $s$, we refit the nuisance functions and
pool the out-of-fold scores across all $L$ folds to obtain
$\hat\theta_s(r)$ and its standard error $\hat s_s(r)$
as in Proposition~\ref{cf:varianceprop}.
Define
\begin{align*}
 \hat\theta_{\rm med}(r)
 &=\operatorname{median}_{1\leq s\leq S}\hat\theta_s(r),\qquad
 \hat s_{\rm med}(r)
 =\operatorname{median}_{1\leq s\leq S}
 \sqrt{\hat s_s(r)^2+
       \{\hat\theta_s(r)-\hat\theta_{\rm med}(r)\}^2}.
\end{align*}

Suppose the preceding assumptions and bandwidth conditions hold for each partition.
Since $S$ is fixed and the leading term in \eqref{asylin}
is common to all partitions, $\hat\theta_{\rm med}(r)$ has
the same first-order expansion.\footnote{For fixed $S$, the common leading term in \eqref{asylin}
implies that $\hat\theta_{\rm med}(r)$ has the same
first-order expansion and
$\max_{s\leq S}|\hat\theta_s(r)-\hat\theta_{\rm med}(r)|
=o_p((nH)^{-1/2})$.
Together with Proposition~\ref{cf:varianceprop}, this
gives consistency of the adjusted standard error
and the stated studentized limit.} Moreover,
 $\sqrt{nH}\,\hat s_{\rm med}(r)
 \xrightarrow{p}\sqrt{V_1(r)}$ and 
$ {(\hat\theta_{\rm med}(r)-\theta(r))}/
      {\hat s_{\rm med}(r)}
 \xrightarrow{d}N(0,1)$.
Thus a pointwise asymptotic $(1-\alpha)$ confidence interval is $ \hat\theta_{\rm med}(r)
 \pm z_{1-\alpha/2}\hat s_{\rm med}(r)$.
Section~\ref{sec:simulation} implements this procedure with $L=5$, $S=11$, and $95\%$ confidence intervals.

Joint inference at finitely many evaluation points is straightforward.
Let $r_1,\ldots,r_J$ be distinct, prespecified interior
evaluation points, with $J$ fixed.
Suppose the conditions of Theorem~\ref{TAsym1} hold
at every $r_j$, using the same nuisance fits and a
common bandwidth $h$.
Write
$\boldsymbol\theta
 =(\theta(r_1),\ldots,\theta(r_J))^\top$ and $\hat{\boldsymbol\theta}
 =(\hat\theta(r_1),\ldots,\hat\theta(r_J))^\top$.
The estimators have the joint limit,  $\sqrt{nh^d}\,
 (\hat{\boldsymbol\theta}-\boldsymbol\theta)
 \xrightarrow{d}N_J(0,\mathbf V)$, 
with $\mathbf V
 =\operatorname{diag}\{V_1(r_1),\ldots,V_1(r_J)\}$.\footnote{
Since $J$ is fixed, the remainders in \eqref{asylin}
are jointly negligible, and the Lyapunov bound in the
proof of Theorem~\ref{TAsym1} applies to every fixed
linear combination of the oracle summands.
For distinct fixed points, their limiting
cross-covariances are zero: the kernel supports are
eventually disjoint for a compactly supported kernel,
and Gaussian kernel overlap vanishes exponentially.}
Let
 $\hat{\boldsymbol\Sigma}_n
 =(nh^d)^{-1}
 \operatorname{diag}\{\hat V_1(r_1),\ldots,\hat V_1(r_J)\}$.
For a fixed full-row-rank matrix
$C\in\mathbb R^{m\times J}$ and a fixed vector
$c\in\mathbb R^m$, Proposition~\ref{cf:varianceprop}
implies that, under $H_0:C\boldsymbol\theta=c$,
\begin{equation}\label{cf:finitewald}
 (C\hat{\boldsymbol\theta}-c)^\top
 (C\hat{\boldsymbol\Sigma}_nC^\top)^{-1}
 (C\hat{\boldsymbol\theta}-c)
 \xrightarrow{d}\chi_m^2.
\end{equation}
The same joint-inference argument applies to the local linear estimators in Theorem~\ref{TAsym23} under their respective conditions.

\subsection{Other conditioning sets and discrete covariates}

\begin{theorem}[Other ICLATEs and fixed discrete conditioning variables]\label{TAsym23}
Let Assumptions~\ref{ass:independence_cov}--\ref{ass:comsup_cov} hold.
Assume Assumption~\ref{ass:support} with $R$ replaced by
$Q$ in part~(i), by $\tilde X$ in part~(ii), and by
$R_c$ under the conditional law given $J_c=1$ in part~(iii).
\begin{enumerate}[label=(\roman*)]
 \item For $\text{ICLATE}_2$, use the index $Q=(\pi(0,X),\pi(1,X))^\top$ and $d=2$. Assume Assumptions~\ref{Akernel}--\ref{Aindex} for this index, redefining $\gamma_\Psi$, $\rho_A$, and its conditional projection accordingly. The preceding expansion and variance formula apply. In particular, $\rho_B=0$ still holds; local continuity for $R$ does not by itself establish local continuity for the coarser index $Q$.
 \item For $\text{ICLATE}_3$ with continuously distributed $\tilde X$, replace $R$ by $\tilde X$ and $d$ by $\dim(\tilde X)$. Assumptions~\ref{Akernel}--\ref{A1stCLATE} suffice; Assumption~\ref{Aindex} is unnecessary because the conditioning variables are observed. The denominator is $E[\pi(1,X)-\pi(0,X)\mid\tilde X=\tilde x]$, which must be positive at the evaluation point.

\item Suppose
$\tilde X=(\tilde X_{\rm con}^\top,\tilde X_{\rm dis}^\top)^\top$,
where $\tilde X_{\rm dis}$ has fixed finite support.
Fix a cell $c$ with
$J_c=1\{\tilde X_{\rm dis}=c\}$ and $p_c=E[J_c]>0$.
For $\text{ICLATE}_1$, let
$R_c=(\tilde X_{\rm con}^\top,\pi(0,X),\pi(1,X))^\top$
and $d_c=\dim(R_c)$.
Retain the cross-fitted scores and use local linear
regression with weights
$J_{c,i}\mathcal K_h(\hat R_{c,i}-r_c)$.
Impose Assumptions~\ref{Akernel}--\ref{Aindex}
under the conditional law given $J_c=1$,
with $R$ and $d$ replaced by $R_c$ and $d_c$.

Write $\gamma_{\Psi,c}(t)=E[\Psi\mid J_c=1,R_c=t]$,
$\theta_c(t)=\gamma_{A,c}(t)/\gamma_{B,c}(t)$,
and $g_c(t)=p_cf_{R_c\mid J_c=1}(t)$.
The preceding expansion and feasible inference apply,
with normalization $\sqrt{nh^{d_c}}$ and variance
 $V_c(r_c)=
 {R_k^{d_c}
 \operatorname{var}(A-\theta_c(r_c)B\mid J_c=1,R_c=r_c)}
 /{(g_c(r_c)\gamma_{B,c}(r_c)^2)}$.

The feasible variance is \eqref{varest} with $d=d_c$,
$\hat\theta(r)$ and $\hat\gamma_B(r)$ replaced by their
within-cell counterparts, and the weights in
\eqref{llweights} recomputed using
$J_{c,i}\mathcal K_h(\hat R_{c,i}-r_c)$
in both the individual weight and the design matrix,
with slopes only in the continuous coordinates.
\end{enumerate}
\end{theorem}

If $\text{ICLATE}_3$ conditions only on a discrete cell, use the ratio of within-cell score sums without a kernel or slope terms. Under the analogous within-cell nuisance conditions with normalization $\sqrt n$, the variance is $\operatorname{var}(A-\theta_c B\mid J_c=1)/(p_c\gamma_{B,c}^2)$. 
With $\hat p_c=n^{-1}\sum_iJ_{c,i}$ and $\hat\gamma_{B,c}=\sum_iJ_{c,i}\hat B_i/\sum_iJ_{c,i}$, a consistent variance estimator is
$\hat V_c={n^{-1}\sum_i J_{c,i}
             (\hat A_i-\hat\theta_c\hat B_i)^2}/
            {\big(\hat p_c^2\hat\gamma_{B,c}^2\big)}$, and the standard error is $\sqrt{\hat V_c/n}$.

	\section{Simulation study}\label{sec:simulation}

We examine the finite-sample performance of
$\widehat{\text{ICLATE}}_1$ under designs with zero and nonzero
index bias. We use sample sizes $n=4000$ and $n=20{,}000$,
with 500 Monte Carlo replications for each design and sample size.

We first describe the data-generating process.
Draw $X\in\mathbb R^{10}$ from $N(0,\Sigma)$ conditional on
$X^\top\Sigma^{-1}X\leq b^2$, where
$\Sigma_{jk}=0.5^{|j-k|}$, $b=2.5$, and $\beta_j=0.5/j$.
Independently of $X$, draw $\upsilon\sim N(0,1)$ and
$(\varepsilon_1,\varepsilon_2)$ jointly standard normal with
correlation $0.5$, with $\upsilon$ independent of this pair.
Generate
\begin{align}\label{dgp}
 Z&=I\{X^\top\beta+\upsilon>0\},\notag\\
 D&=I\{-0.3+X^\top\beta+\gamma(X)Z+\varepsilon_1>0\},
 \qquad \gamma(X)=0.9+0.25X_2,\notag\\
 Y&=X^\top\beta+D(X_1+\lambda U+\varepsilon_2)+\varepsilon_2,
 \qquad \lambda\in\{0,1\},
\end{align}
where $I\{\cdot\}$ is the indicator function.
The covariate of interest is $\tilde X=X_1$.
The positive shift $\gamma(X)$ ensures monotonicity, while
correlation between $\varepsilon_1$ and $\varepsilon_2$
generates selection on unobservables.

Let $\xi(X)=(X_1,X_2,X^\top\beta)^\top$ and
$U=\{X_3-a_U^\top\xi(X)\}/s_U$, where $a_U$ and $s_U$
are the population linear-projection coefficients and 
standard deviation of the population linear-projection error, under the truncated covariate distribution.
These constants, not $U$, are fixed across replications.
Appendix~\ref{app:simcontinuity} gives their formulas and population values and shows that
$E[U\mid R]=0$ for $R=(X_1,\pi(0,X),\pi(1,X))^\top$.
Consequently,
\[
 \rho_A(X)=\lambda\{\pi(1,X)-\pi(0,X)\}U,
\]
so $\lambda=0$ and $\lambda=1$ give zero and nonzero index bias,
respectively, but the same population value of $\text{ICLATE}_1$:
\begin{equation}\label{truth}
 \Delta(\tilde x,c,\pi',\pi'')
 =\tilde x+\frac{
 \phi\{-\Phi^{-1}(\pi'')\}-\phi\{-\Phi^{-1}(\pi')\}
 }{2(\pi''-\pi')},
\end{equation}
where $\Phi$ and $\phi$ are the standard normal cdf and density.
We evaluate at $(\pi',\pi'')=(0.40,0.70)$ and
$\tilde x\in\{-0.5,0,0.5\}$, giving true effects of
approximately $-0.564$, $-0.064$, and $0.436$ in both designs.
Appendix~\ref{app:simcontinuity} collects the design diagnostics and
verifies the support, regularity, nuisance-rate, and index conditions
in Assumptions~\ref{ass:support} and~\ref{Akernel}--\ref{Aindex},
under the stated regularity for the parametric nuisance estimators.

Estimation follows Sections~\ref{sec:estimation}
and~\ref{sec:asymptotics}.
We estimate $p(X)$ by probit regression on $X$, and estimate
$\pi(z,X)$ by separate probits within each instrument group.
Within each group, we estimate $\mu(z,X)$ by linear regression
on an intercept, $X$, $\hat\pi(z,X)$,
$X\hat\pi(z,X)$, and $\phi(\hat t_z(X))$, where
$\hat t_z$ is the fitted probit index.
These specifications contain the true nuisance functions;
their population formulas are given in
Appendix~\ref{app:simcontinuity}.
All nuisance fits use only the relevant training sample.
We clip fitted instrument propensities to $[0.01,0.99]$ and
fitted treatment propensities to $[10^{-4},1-10^{-4}]$,
retaining all observations and using the scores as defined in
Section~\ref{sec:estimation}.

The second step uses pooled local linear regression with a
Gaussian product kernel and the same bandwidth vector for the
numerator and denominator. We set
\[
 \hat h_j=1.06\,\hat\sigma_j n^{-1.2/(d_R+4)},
 \qquad j=1,\ldots,d_R,\qquad d_R=3,
\]
where $\hat\sigma_j$ is the sample standard deviation of the
$j$th fitted conditioning variable.
We use five-fold cross-fitting, repeat the procedure over
$S=11$ partitions, and report the median estimate and adjusted
standard error defined in Section~\ref{sec:asymptotics}.
Partition-specific variances use \eqref{varest}, and coverage
is evaluated for
$\hat\theta_{\rm med}(r)\pm1.96\hat s_{\rm med}(r)$.

Table~\ref{tab:sim} reports absolute bias, standard deviation,
RMSE, mean estimated standard error, and coverage.
Absolute bias is below $0.09$ and below $11\%$ of the empirical
standard deviation in every configuration, with RMSE close to
the standard deviation.
Increasing $n$ from $4000$ to $20{,}000$ reduces standard
deviations to approximately $0.55$--$0.72$ of their earlier
values, broadly consistent with the factor
$5^{-17/70}\approx0.68$ implied by Theorem~\ref{TAsym1}
and the chosen bandwidth.

\begin{table}[ht]
		\small
		\centering
		\caption{Finite sample behavior of $\widehat{\text{ICLATE}}_1$ at $\pi'=0.40$, $\pi''=0.70$}
		\label{tab:sim}
		\begin{tabular}{lrccccc}
			\toprule
			& True value & $|$Bias$|$ & Std.\ dev. & RMSE & Mean s.e. & Coverage \\
			\midrule
			\multicolumn{7}{l}{\textit{$\lambda=0$, no index bias}}\\
			\multicolumn{7}{l}{\quad\textit{$n=4000$}}\\
			$\tilde x=-0.5$ & $-0.564$ & 0.012 & 0.978 & 0.977 & 1.088 & 0.970 \\
			$\tilde x=0$ & $-0.064$ & 0.007 & 0.633 & 0.633 & 0.688 & 0.974 \\
			$\tilde x=0.5$ & $0.436$ & 0.030 & 0.957 & 0.956 & 0.989 & 0.980 \\
			\multicolumn{7}{l}{\quad\textit{$n=20{,}000$}}\\
			$\tilde x=-0.5$ & $-0.564$ & 0.074 & 0.701 & 0.704 & 0.753 & 0.976 \\
			$\tilde x=0$ & $-0.064$ & 0.002 & 0.384 & 0.384 & 0.444 & 0.978 \\
			$\tilde x=0.5$ & $0.436$ & 0.021 & 0.531 & 0.531 & 0.541 & 0.962 \\[4pt]
			\multicolumn{7}{l}{\textit{$\lambda=1$, nonzero index bias}}\\
			\multicolumn{7}{l}{\quad\textit{$n=4000$}}\\
			$\tilde x=-0.5$ & $-0.564$ & 0.031 & 1.075 & 1.075 & 1.136 & 0.970 \\
			$\tilde x=0$ & $-0.064$ & 0.010 & 0.667 & 0.666 & 0.720 & 0.966 \\
			$\tilde x=0.5$ & $0.436$ & 0.057 & 1.003 & 1.003 & 1.037 & 0.976 \\
			\multicolumn{7}{l}{\quad\textit{$n=20{,}000$}}\\
			$\tilde x=-0.5$ & $-0.564$ & 0.082 & 0.768 & 0.772 & 0.785 & 0.958 \\
			$\tilde x=0$ & $-0.064$ & 0.002 & 0.424 & 0.424 & 0.466 & 0.974 \\
			$\tilde x=0.5$ & $0.436$ & 0.032 & 0.568 & 0.568 & 0.567 & 0.948 \\
			\bottomrule
		\end{tabular}
		\par\smallskip
\begin{minipage}{\textwidth}\footnotesize\textit{Notes:}
Results are based on 500 replications per design and sample size.
Estimates and standard errors use the repeated-partition procedure
with $L=5$ and $S=11$.
``Mean s.e.'' averages the adjusted standard errors across replications.
Coverage is the fraction of the reported $95\%$ confidence intervals
containing the true value in \eqref{truth}.
\end{minipage}
\end{table}

Mean estimated standard errors are generally larger than the
empirical standard deviations, by up to approximately $16\%$.
Coverage ranges from $0.948$ to $0.980$, with most values above
$0.95$; the results do not isolate the contribution of the
repeated-partition adjustment to this overcoverage.
The larger standard deviations at $\lambda=1$ also reflect
increased oracle score variance, not solely estimation of the
conditioning index. Isolating the latter requires a comparison
with an estimator using the true index within each design.

\section{Application to the Job Corps program}\label{sec:application}

	We illustrate the approach with the National Job Corps Study, an experimental evaluation of a training program for disadvantaged youth in the United States, in which applicants aged 16 to 24 were randomly assigned to eligibility for the program in the mid 1990s; see \citet{Schochet_Burghardt_McConnell_2008} for the original impact findings. The data have been used extensively in the treatment effect literature, for instance by \citet{Lee_2009} for bounds under sample selection, \citet{Flores_etal_2012} for the effect of the length of exposure to instruction, and \citet{Frolich_Huber_2017} for mediation analysis with instruments. We use the version of the data supplied with the \texttt{causalweight} package for \texttt{R} of \citet{Bodory_Huber_2018}.

	The design maps into our framework as follows. The instrument $Z$ is random assignment to eligibility. The treatment $D$ indicates participation in academic or vocational training during the first year after assignment. 
     Random assignment supports instrument exogeneity.
Exclusion and monotonicity for the specified training
indicator are maintained assumptions.
    The outcome $Y$ is weekly earnings in the fourth year after assignment, in US dollars. The covariates $X$ are the 28 characteristics recorded at assignment, including gender, age, ethnicity, education and degrees, household composition, parental education, prior employment and earnings, and health and behavioral measures. The covariate of interest $\tilde{X}$ is age at assignment, which we treat as quasi-continuous.

	No observation has an estimated instrument propensity score outside $[0.01,0.99]$, so the clipping of Section~\ref{sec:simulation} leaves all 9,240 observations unaltered. This reflects that assignment was randomized, so that the instrument propensity score is far from the boundaries throughout. Non-compliance is two-sided: a substantial share of those assigned to eligibility do not enroll, and some of those not assigned obtain training elsewhere. The complier population may vary with the covariates.   Estimation follows Section~\ref{sec:simulation}, with the nuisance functions estimated by linear and probit regressions on the covariates, the two treatment propensity scores being estimated separately on the subsamples with $Z=1$ and $Z=0$, with five-fold cross-fitting repeated over eleven splits, clipping as described there, and a local linear second step with a Gaussian product kernel and an undersmoothed rule-of-thumb bandwidth. Standard errors are computed from \eqref{varest}.
         The outcome regressions are not augmented, since the
augmentation in Section~\ref{sec:simulation} is specific
to its DGP.
We use linear outcome regressions and probit propensity
models as working specifications.
The inference results require the nuisance-accuracy
and generated-index conditions in
Assumptions~\ref{A1stCLATE} and~\ref{Aindex}. 

\begin{figure}[ht]
		\centering
		\caption{Joint support of the fitted treatment propensity scores}
		\label{fig:jcsupport}
		\includegraphics[width=0.7\textwidth]{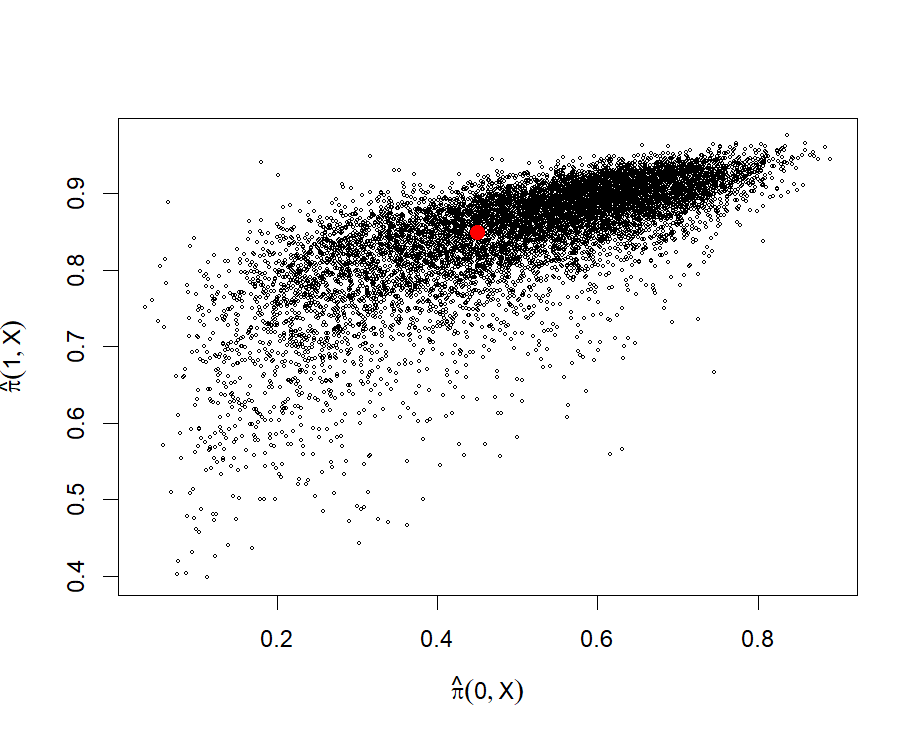}
		\par\smallskip
		\begin{minipage}{0.9\textwidth}
			\footnotesize\textit{Notes:} Cross-fitted estimates of $\pi(0,X)$ and $\pi(1,X)$ for the estimation sample, obtained from probit regressions with five-fold cross-fitting. The red point marks the evaluation point $(\pi',\pi'')=(0.45,0.85)$ used in Tables~\ref{tab:jc1} and \ref{tab:jc2}. 
            For the age-specific estimates, the relevant support
condition must hold conditional on each reported age. 
		\end{minipage}
	\end{figure}

	Before choosing an evaluation point we inspect the joint support of the two fitted propensity scores, since Assumption \ref{ass:support} restricts which pairs are admissible. The two scores trace out a genuinely two-dimensional region rather than a curve, with $\hat\pi(0,X)$ spread over roughly $0.05$ to $0.9$ and $\hat\pi(1,X)$ lying mostly between $0.75$ and $0.95$ with a thinner tail reaching down to about $0.4$.  
Figure~\ref{fig:jcsupport} summarizes the joint variation
in the fitted treatment propensity scores and informs
the choice of the evaluation pair.
For the age-specific comparisons, we maintain the
local-support condition in Assumption~\ref{ass:support}
at each reported age.
The effective local sample sizes in Table~\ref{tab:jc1}
provide a complementary diagnostic of kernel-weight
concentration at these evaluation points.
 Complier share and local density pull against each other along that region, since the largest values of $\pi''-\pi'$ occur where $\hat\pi(0,X)$ is small and the data are thinner. The attainable complier shares have a median of $0.34$ and an interquartile range of $0.26$ to $0.43$. Considering this trade-off, we ultimately select $\pi'=0.45$ and $\pi''=0.85$, implying a complier share of $0.40$.

	\begin{table}[ht]
		\small
		\centering
		\caption{$\text{ICLATE}_1$ across age at assignment at $\pi'=0.45$, $\pi''=0.85$}
		\label{tab:jc1}
		\begin{tabular}{lcccc}
			\toprule
			Age at assignment & $\widehat{\text{ICLATE}}_1$ & Std.\ error & $p$-value & $n_{\text{eff}}$ \\
			\midrule
			17 & {111.2} & {83.6} & {0.184} & {241} \\
			18 & { 16.0} & {54.3} & {0.768} & {417} \\
			19 & { 33.5} & {54.7} & {0.540} & {413} \\
			20 & {198.7} & {74.1} & {0.007} & {236} \\
			\bottomrule
		\end{tabular}
		\par\smallskip
		\begin{minipage}{\textwidth}\footnotesize\textit{Notes:} Outcome is weekly earnings in the fourth year after assignment, in US dollars. Nuisance functions estimated by linear and {probit} regressions with {five-fold} cross-fitting repeated over eleven splits, {the fitted instrument propensity score clipped to $[0.01,0.99]$ and the fitted treatment propensity scores to $[10^{-4},1-10^{-4}]$}; second step by local linear regression with a Gaussian product kernel and an undersmoothed rule-of-thumb bandwidth. Estimates are medians across splits. $n_{\text{eff}}=\left(\sum_i w_i\right)^2/\sum_i w_i^2$ is the effective local sample size at the evaluation point, with $w_i$ the kernel weights.\end{minipage}
	\end{table}

	Table~\ref{tab:jc1} reports $\text{ICLATE}_1$ across age at assignment, holding the range of unobservables fixed. All four estimates are positive, ranging from $16$ to $199$ dollars of weekly earnings. Only the effect at age 20 is significant, at the 1 percent level. The point estimate is largest at age 20, but the
reported pointwise tests concern zero effects,
not differences across ages. The effective local sample sizes are between 236 and 417 observations across the four ages.

	Table~\ref{tab:jc2} turns to $\text{ICLATE}_2$, which holds the pair of propensity scores fixed while averaging over the covariates, and reports it across values of $\pi'$. This is the comparison that Proposition~\ref{prop:test} concerns, in the $\text{ICLATE}_2$ version discussed at the end of Section~\ref{sec:extrapolation}. Relative to Table~\ref{tab:jc1} it lowers $d_R$ from three to two, at the price that moving $\pi'$ shifts the composition of the entire covariate vector, age included, and not only of the covariates excluded from $\tilde{X}$. Raising $\pi'$ from $0.35$ to $0.55$ at a fixed $\pi''=0.85$ narrows the complier interval toward its upper end, that is, toward individuals whose participation requires a high treatment propensity, and shifts that composition at the same time.
	 The estimates range from $68.4$ to $96.3$ dollars.
The reported $p$-values test whether each effect is zero, and one of
the three is significant at the 10 percent level.

	\begin{table}[ht]
		\small
		\centering
		\caption{$\text{ICLATE}_2$ across propensity score pairs}
		\label{tab:jc2}
		\begin{tabular}{ccccccc}
			\toprule
			$\pi'$ & $\pi''$ & $\pi''-\pi'$ & $\widehat{\text{ICLATE}}_2$ & Std.\ error & $p$-value & $n_{\text{eff}}$ \\
			\midrule
			0.35 & 0.85 & 0.50 & {68.4} & {53.4} & {0.200} & {404} \\
			0.45 & 0.85 & 0.40 & {73.9} & {39.3} & {0.060} & {665} \\
			0.55 & 0.85 & 0.30 & {96.3} & {61.2} & {0.116} & {667} \\
			\bottomrule
		\end{tabular}
		\par\smallskip
		\begin{minipage}{\textwidth}\footnotesize\textit{Notes:} As in Table~\ref{tab:jc1}. $\text{ICLATE}_2$ conditions on the two propensity scores only, so that $d_R=2$ and the covariates are averaged over. The grid runs over $\pi'$ rather than $\pi''$ because $\hat\pi(1,X)$ is concentrated around $0.85$ whereas $\hat\pi(0,X)$ is spread out, so that moving $\pi'$ sweeps a wider range of the unobservable at comparable local density.\end{minipage}
	\end{table}

	Table~\ref{tab:jc3} reports $\text{ICLATE}_3$, the complier effect conditional on age alone. It is estimated far more precisely, with standard errors roughly a third to a half of those of Table~\ref{tab:jc1}, because the conditioning set is one-dimensional rather than three-dimensional and the effective local sample sizes are substantially larger. The estimates are positive throughout and range from $14$ to $95$ dollars, with the effect at age 20 significant at the 1 percent level and that at age 17 at the 10 percent level. Overall, the age-specific estimates are positive, with
statistically significant effects at age 20 for both
the first and third ICLATEs.
The reported pointwise results do not establish
differences across ages.

	\begin{table}[ht]
		\small
		\centering
		\caption{$\text{ICLATE}_3$ across age at assignment}
		\label{tab:jc3}
		\begin{tabular}{lcccc}
			\toprule
			Age at assignment & $\widehat{\text{ICLATE}}_3$ & Std.\ error & $p$-value & $n_{\text{eff}}$ \\
			\midrule
			17 & {58.4} & {32.0} & {0.068} & {1672} \\
			18 & {33.3} & {27.0} & {0.218} & {1653} \\
			19 & {13.9} & {26.7} & {0.604} & {1297} \\
			20 & {94.7} & {28.0} & {0.001} & { 909} \\
			\bottomrule
		\end{tabular}
		\par\smallskip
		\begin{minipage}{\textwidth}\footnotesize\textit{Notes:} As in Table~\ref{tab:jc1}. $\text{ICLATE}_3$ conditions on age at assignment only, so that $d_R=1$ and no generated regressor enters the second step.\end{minipage}
	\end{table}
	\section{Conclusion}\label{con}

	Comparing local average treatment effects across covariate values is ambiguous, because the complier population generally changes with the covariates, so apparent heterogeneity may reflect the covariates, the unobservables, or both. We have proposed conditioning on the two treatment propensity scores in addition to a subset of covariates of interest. Under the conventional IV assumptions those scores are the quantiles of the unobserved term that delimit the compliers, so that fixing them fixes the range of unobservables and removes that channel from the comparison. The resulting informative conditional LATEs are identified by a ratio of conditional expectations of doubly robust scores, under the conditional LATE assumptions together with a support condition: the covariates that are controlled for but not reported on must generate variation in the propensity scores. 
    	The three parameters differ in what they hold fixed,
implying different comparison groups.
$\text{ICLATE}_1$ holds both the covariates of interest
and the range of the normalized unobservable fixed and
supports comparisons across those covariates;
$\text{ICLATE}_2$ holds only that range fixed and
supports comparisons across ranges of the unobservable
while averaging over covariates;
$\text{ICLATE}_3$ holds only the covariates of interest fixed.
	
	 We distinguish effect homogeneity in the unobservable from homogeneity across excluded covariates. Together, these restrictions provide sufficient conditions for the first and third ICLATEs to equal the average
treatment effect conditional on the covariates of interest.
Their joint validity implies constancy across
propensity-score pairs, while homogeneity across excluded
covariates also implies interval additivity (meaning that the effect over the union of two adjacent propensity score intervals is an interval length-weighted average effect) under the
required support conditions.
Under the maintained IV assumptions, rejection of
constancy provides evidence against the joint restrictions;
maintaining either one attributes the rejection to the other. 

For estimation, we combine cross-fitted nuisance estimation
with pooled local linear regression of doubly robust scores.
Under the stated accuracy, continuity, and bandwidth
conditions, we establish feasible pointwise inference
and joint inference at finitely many prespecified
evaluation points.
The simulations assess the finite-sample performance of the implemented estimator and inference procedure.
In the reported simulation designs, bias is small relative to the standard deviation, and confidence
interval coverage is generally close to or above the nominal level.
 In the Job Corps application, the age-specific estimates
are positive, with statistically significant effects
at age 20 both at the selected propensity-score pair
and when averaging over propensity-score pairs.
The reported pointwise results do not establish
statistically significant differences across ages.

\section*{Declaration of AI use}
During the preparation of this work, the authors used ChatGPT (OpenAI) and Claude (Anthropic) to assist in language editing, mathematical derivations and proof checking, and reference verification. The authors take full responsibility for reviewing and verifying all mathematical arguments, references, and other content in the final manuscript.

    \clearpage
	\bibliographystyle{elsarticle-harv}
	\bibliography{BIB_MZ}

@ARTICLE{Angrist+96,
 author    = {J. Angrist and G. Imbens and D. Rubin},
 title   = {Identification of Causal Effects using Instrumental Variables},
 year      = {1996},
 journal   =  {Journal of American Statistical Association},
 volume    = {91},
 pages     = {444-472}
}

@Article{Rubin_1974,
author = {Donald B. Rubin},
title = {Estimating Causal Effects of Treatments in Randomized and Nonrandomized Studies},
journal = {Journal of Educational Psychology},
year = {1974},
volume = {66},
number = {5},
pages = {688-701},
}

@article{van_de_Geer_2008,
  author  = {van de Geer, Sara A.},
  title   = {High-dimensional generalized linear models
             and the lasso},
  journal = {The Annals of Statistics},
  year    = {2008},
  volume  = {36},
  number  = {2},
  pages   = {614--645},
}

@article{Belloni_Chernozhukov_Wei_2016,
  author  = {Belloni, Alexandre and Chernozhukov, Victor
             and Wei, Ying},
  title   = {Post-Selection Inference for Generalized
             Linear Models With Many Controls},
  journal = {Journal of Business \& Economic Statistics},
  year    = {2016},
  volume  = {34},
  number  = {4},
  pages   = {606--619},
  doi     = {10.1080/07350015.2016.1166116}
}

@article{Zhao_Ding_Li_2026,
  author  = {Zhao, Anqi and Ding, Peng and Li, Fan},
  title   = {Two-stage least squares with treatment-covariate interactions
             for treatment effect heterogeneity},
  journal = {Journal of Econometrics},
  year    = {2026},
  volume  = {258},
  pages   = {106324},
}

@article{Chernozhukov_Chetverikov_Demirer_Duflo_Hansen_Newey_2017,
author = {Chernozhukov, Victor and Denis Chetverikov and Mert Demirer and Esther Duflo and Christian
Hansen and Whitney Newey and James Robins},
title = {Double/Debiased Machine Learning for Treatment and Structural Parameters},
journal={The Econometrics Journal},
year = {2018},
volume = {21},
number = {1},
pages = {C1-C68},
}

@Article{Robins_Rotnitzky_Zhao_1994,
author = {Robins, James M. and Andrea Rotnitzky and Ping Zhao},
title = {Estimation of Regression Coefficients When Some Regressors Are Not Always Observed},
journal = {Journal of the American Statistical Association},
year = {1994},
volume = {89},
number = {427},
pages = {846-866},
}

@Article{Mammen_Rothe_Schienle_2012,
author = {Mammen, Enno and Christoph Rothe and Melanie Schienle},
title = {Nonparametric Regression with Nonparametrically Generated Covariates},
journal = {The Annals of Statistics},
year = {2012},
volume = {40},
number = {2},
pages = {1132-1170},
}

@Article{MRS2016,
author={Enno Mammen and Christoph Rothe and Melanie Schienle},
year={2016}, 
title={Semiparametric Estimation with Generated Covariates}, 
journal={Econometric Theory},
volume={5},
number={32},
pages={1140-1177}
}

@TechReport{LeePMGR,
author={Ying-Ying Lee},
year={2018},
title={Partial Mean Processes with Generated Regressors: Continuous Treatment Effects and Nonseparable Models},
type={arXiv:1811.00157}
}

@article{CL,
	author={Kyle Colangelo and Ying-Ying Lee},
	title={Double Debiased Machine Learning Nonparametric Inference with
	Continuous Treatments},
	journal={Journal of Business \& Economic Statistics},
	year={2026},
pages={67-79},
VOLUME={44},
NUMBER={1}
}

@Article{Hahn_Ridder_2013,
author = {Hahn, Jinyong and Geert Ridder},
title = {Asymptotic Variance of Semiparametric Estimators With Generated Regressors},
journal = {Econometrica},
year = {2013},
volume = {81},
number = {1},
pages = {315-340},
}

@Article{Abrevaya_Hsu_Lieli_2015,
author = {Abrevaya, Jason and Yu-Chin Hsu and Robert P. Lieli},
title = {Estimating Conditional Average Treatment Effects},
journal = {Journal of Business and Economic Statistics},
year = {2015},
volume = {33},
number = {4},
pages = {485-505},
}

@Article{Wager_Athey_2018,
author = {Wager, Stefan and Susan Athey},
title = {Estimation and Inference of Heterogeneous Treatment Effects using Random Forests},
journal = {Journal of the American Statistical Association},
year = {2018},
volume = {113},
number = {523},
pages = {1228-1242},
}

@Article{Lee_Okui_Whang_2016,
author = {Lee, Sokbae and Ryo Okui and Yoon-Jae Whang},
title = {Doubly robust uniform confidence band for the conditional average treatment effect function},
journal = {Journal of Applied Econometrics},
year = {2017},
volume = {32},
number = {7},
pages = {1207-1225},
}

@Article{Athey_Tibshirani_Wager_2019,
author = {Athey, Susan and Julie Tibshirani and Stefan Wager},
title = {Generalized random forests},
journal = {The Annals of Statistics},
year = {2019},
volume = {47},
number = {2},
pages = {1148-1178},
}

@online{Lechner_2018,
author = {Lechner, Michael},
title = {Modified Causal Forests for Estimating Heterogeneous Causal Effects},
year = {2018},
version = {2},
eprinttype = {arxiv},
eprint = {1812.09487v2},
}

@Article{Semenova_Chernozhukov_2021,
author = {Semenova, Vira and Victor Chernozhukov},
title = {Debiased Machine Learning of Conditional Average Treatment Effects and Other Causal Functions},
journal = {The Econometrics Journal},
year = {2021},
volume = {24},
number = {2},
pages = {264-289},
}

@Article{Fan_Hsu_Lieli_Zhang_2019,
author = {Fan, Qingliang and Yu-Chin Hsu and Robert P. Lieli and Yichong Zhang},
title = {Estimation of Conditional Average Treatment Effects with High-Dimensional Data},
journal = {Journal of Business and Economic Statistics},
year = {2022},
volume = {40},
number = {1},
pages = {313-327},
}

@Article{Imbens_Angrist_1994,
 ISSN = {00129682, 14680262},
 author = {Guido W. Imbens and Joshua D. Angrist},
 journal = {Econometrica},
 number = {2},
 pages = {467--475},
 publisher = {[Wiley, Econometric Society]},
 title = {Identification and Estimation of Local Average Treatment Effects},
 volume = {62},
 year = {1994}
}

@Article{Vytlacil_2002,
  title={Independence, monotonicity, and latent index models: An equivalence result},
  author={Vytlacil, Edward},
  journal={Econometrica},
  volume={70},
  number={1},
  pages={331--341},
  year={2002},
  publisher={Wiley Online Library}
}

@article{Heckman_Vytlacil_1999,
  title={Local instrumental variables and latent variable models for identifying and bounding treatment effects},
  author={Heckman, James J and Vytlacil, Edward J},
  journal={Proceedings of the national Academy of Sciences},
  volume={96},
  number={8},
  pages={4730--4734},
  year={1999},
  publisher={National Acad Sciences}
}

@article{Heckman_Vytlacil_2005,
  title={Structural equations, treatment effects, and econometric policy evaluation 1},
  author={Heckman, James J and Vytlacil, Edward},
  journal={Econometrica},
  volume={73},
  number={3},
  pages={669--738},
  year={2005},
  publisher={Wiley Online Library}
}

@article{Heckman_Urzua_Vytlacil_2006,
  title={Understanding instrumental variables in models with essential heterogeneity},
  author={Heckman, James J and Urzua, Sergio and Vytlacil, Edward},
  journal={The Review of Economics and Statistics},
  volume={88},
  number={3},
  pages={389--432},
  year={2006},
  publisher={MIT Press}
}

@article{Abadie_2003,
  title={Semiparametric instrumental variable estimation of treatment response models},
  author={Abadie, Alberto},
  journal={Journal of econometrics},
  volume={113},
  number={2},
  pages={231--263},
  year={2003},
  publisher={Elsevier}
}

@article{Frolich_2007,
  title={Nonparametric IV estimation of local average treatment effects with covariates},
  author={Fr{\"o}lich, Markus},
  journal={Journal of Econometrics},
  volume={139},
  number={1},
  pages={35--75},
  year={2007},
  publisher={Elsevier}
}

@online{Zimmert_Lechner_2019,
  title={Nonparametric estimation of causal heterogeneity under high-dimensional confounding},
  author={Zimmert, Michael and Lechner, Michael},
  date = {2019-08-23},
  version = {1},
  eprinttype = {arxiv},
  eprint = {arXiv:1908.08779v1},
  year={2019}
}

@InCollection{Angrist_FernandezVal_2013,
author = {Angrist, Joshua D. and Iv\'{a}n Fern\'{a}ndez-Val},
title = {ExtrapoLATE-ing: External Validity and Overidentification in the LATE Framework},
booktitle = {Advances in Economics and Econometrics: Tenth World Congress, Volume 3},
editor = {Acemoglu, Daron and Manuel Arellano and Eddie Dekel},
series = {Econometric Society Monographs},
publisher = {Cambridge University Press},
address = {Cambridge},
year = {2013},
pages = {401-434},
}

@Article{Aronow_Carnegie_2013,
author = {Aronow, Peter M. and Allison Carnegie},
title = {Beyond LATE: Estimation of the Average Treatment Effect with an Instrumental Variable},
journal = {Political Analysis},
year = {2013},
volume = {21},
number = {4},
pages = {492-506},
}

@Article{Ogburn_Rotnitzky_Robins_2015,
author = {Ogburn, Elizabeth L. and Andrea Rotnitzky and James M. Robins},
title = {Doubly Robust Estimation of the Local Average Treatment Effect Curve},
journal = {Journal of the Royal Statistical Society, Series B},
year = {2015},
volume = {77},
number = {2},
pages = {373-396},
}

@Article{Wang_TchetgenTchetgen_2018,
author = {Wang, Linbo and Tchetgen Tchetgen, Eric},
title = {Bounded, Efficient and Multiply Robust Estimation of Average Treatment Effects Using Instrumental Variables},
journal = {Journal of the Royal Statistical Society, Series B},
year = {2018},
volume = {80},
number = {3},
pages = {531-550},
}

@Article{Brinch_Mogstad_Wiswall_2017,
author = {Brinch, Christian N. and Magne Mogstad and Matthew Wiswall},
title = {Beyond LATE with a Discrete Instrument},
journal = {Journal of Political Economy},
year = {2017},
volume = {125},
number = {4},
pages = {985-1039},
}

@Article{Bearth_Lechner_2025,
author = {Bearth, Nora and Michael Lechner},
title = {Causal Machine Learning for Moderation Effects},
journal = {Journal of Business and Economic Statistics},
year = {2025},
volume = {43},
number = {4},
pages = {1158-1169},
}

@online{Escanciano_PerezIzquierdo_2026,
author = {Escanciano, Juan Carlos and Telmo P\'{e}rez-Izquierdo},
title = {Automatic Locally Robust GMM with Machine-Learning-Generated Regressors},
year = {2026},
note = {arXiv:2301.10643},
}

@Article{Tan_2006,
author = {Tan, Zhiqiang},
title = {Regression and Weighting Methods for Causal Inference Using Instrumental Variables},
journal = {Journal of the American Statistical Association},
year = {2006},
volume = {101},
number = {476},
pages = {1607-1618},
}

@InCollection{Card_1995,
author = {Card, David},
title = {Using Geographic Variation in College Proximity to Estimate the Return to Schooling},
booktitle = {Aspects of Labour Market Behaviour: Essays in Honour of John Vanderkamp},
editor = {Christofides, Louis N. and E. Kenneth Grant and Robert Swidinsky},
publisher = {University of Toronto Press},
address = {Toronto},
year = {1995},
pages = {201-222},
}

@InProceedings{Bodory_Huber_2018,
author = {Bodory, Hugo and Martin Huber},
title = {The causalweight Package for Causal Inference in {R}},
booktitle = {SES Working Paper 493},
publisher = {University of Fribourg},
address = {Fribourg},
year = {2018},
}

@Article{Schochet_Burghardt_McConnell_2008,
author = {Schochet, Peter Z. and John Burghardt and Sheena McConnell},
title = {Does Job Corps Work? Impact Findings from the National Job Corps Study},
journal = {American Economic Review},
year = {2008},
volume = {98},
number = {5},
pages = {1864-1886},
}

@Article{Lee_2009,
author = {Lee, David S.},
title = {Training, Wages, and Sample Selection: Estimating Sharp Bounds on Treatment Effects},
journal = {The Review of Economic Studies},
year = {2009},
volume = {76},
number = {3},
pages = {1071-1102},
}

@Article{Flores_etal_2012,
author = {Flores, Carlos A. and Alfonso Flores-Lagunes and Arturo Gonzalez and Todd C. Neumann},
title = {Estimating the Effects of Length of Exposure to Instruction in a Training Program: The Case of Job Corps},
journal = {The Review of Economics and Statistics},
year = {2012},
volume = {94},
number = {1},
pages = {153-171},
}

@Article{Frolich_Huber_2017,
author = {Fr\"{o}lich, Markus and Martin Huber},
title = {Direct and Indirect Treatment Effects: Causal Chains and Mediation Analysis with Instrumental Variables},
journal = {Journal of the Royal Statistical Society, Series B},
year = {2017},
volume = {79},
number = {5},
pages = {1645-1666},
}

@InCollection{Mogstad_Torgovitsky_2024,
author = {Mogstad, Magne and Torgovitsky, Alexander},
title = {Instrumental Variables with Unobserved Heterogeneity in Treatment Effects},
booktitle = {Handbook of Labor Economics},
publisher = {Elsevier},
year = {2024},
volume = {5},
pages = {1-114}
}
	
	\bigskip
	
	\renewcommand\appendix{\par
		\setcounter{section}{0}%
		\setcounter{subsection}{0}%
		\setcounter{table}{0}%
		\setcounter{figure}{0}%
		\renewcommand\thesection{\Alph{section}}%
		\renewcommand\thetable{\Alph{section}.\arabic{table}}}
	\renewcommand\thefigure{\Alph{section}.\arabic{subsection}.\arabic{subsubsection}.\arabic{figure}}
	\clearpage
	
	\begin{appendix}
		
		\numberwithin{equation}{section}
		\noindent \textbf{\LARGE Appendices}
		
		\section{Proofs of the identification results}\label{appI}
		
		\paragraph{Proof of Theorem~\ref{thm:ident}}
		Substituting \eqref{condLATE} for $\Delta(X,c)$ in \eqref{firsteff2} and writing $\mathcal{A}=\{\tilde{X}=\tilde{x},Q=q\}$ for brevity,
		\begin{eqnarray}\label{firsteff3proof}
			&&E\left[\frac{\mu(1,X)-\mu(0,X)}{\pi(1,X)-\pi(0,X)}\Big|\mathcal{A}, T=c \right]\notag\\
			&=&E\left[\frac{\mu(1,X)-\mu(0,X)}{\Pr(T=c|X)}\cdot\frac{\Pr(T=c|X,\mathcal{A})}{\Pr(T=c|\mathcal{A})}   \Big| \mathcal{A}\right]\notag\\
			&=&E\left[\frac{\mu(1,X)-\mu(0,X)}{\Pr(T=c|X)}\cdot\frac{\Pr(T=c|X)}{\Pr(T=c|\mathcal{A})}   \Big| \mathcal{A}\right]\notag\\
			&=&\frac{E\left[\mu(1,X)-\mu(0,X)\big| \mathcal{A}\right]}{\Pr(T=c|\mathcal{A})}
			\ =\ \frac{E\left[\mu(1,X)-\mu(0,X)\big| \mathcal{A}\right]}{E\left[\pi(1,X)-\pi(0,X)\big| \mathcal{A}\right]}.
		\end{eqnarray}
		The first equality follows from Bayes' theorem together with $\pi(1,X)-\pi(0,X)=\Pr(T=c|X)$, which holds under Assumptions \ref{ass:independence_cov} and \ref{ass:monotonicity_cov}. The second equality uses $\tilde{X}\subseteq X$ and the fact that $\pi(0,X)$ and $\pi(1,X)$ are deterministic functions of $X$, so that the event $\mathcal{A}$ is measurable with respect to $\sigma(X)$ and hence $\Pr(T=c|X,\mathcal{A})=\Pr(T=c|X)$. The third equality holds because $\Pr(T=c|\mathcal{A})$ does not depend on $X$, and the fourth follows from iterated expectations and, again, $\pi(1,X)-\pi(0,X)=\Pr(T=c|X)$. Assumptions \ref{ass:first_stage_cov} and \ref{ass:support} guarantee that the denominator is strictly positive and that all conditional expectations are well defined. The arguments for $\text{ICLATE}_2$ and $\text{ICLATE}_3$ are identical upon replacing $\mathcal{A}$ by the corresponding conditioning event.
		\hfill $\square$
		
		\paragraph{Proof of Proposition~\ref{thm:mte}}
		Under Assumption \ref{ass:independence_cov}, $\bar V$ is uniformly distributed on $[0,1]$ and independent of $(Z,X)$, and hence of $R=r_0(X)$. Conditional on $R=r$, the propensity scores are fixed at $(\pi',\pi'')$, so that the compliance event $\{T=c\}$ coincides with $\{\bar V\in(\pi',\pi'']\}$, an event in $\sigma(\bar V)$ alone. Therefore, conditional on $R=r$ and $T=c$, $\bar V$ is uniform on $(\pi',\pi'']$ and remains independent of $X$. Iterated expectations then give
		\begin{eqnarray*}
			\text{ICLATE}_1=E\left[E\left[Y(1)-Y(0)\big|X,\bar V\right]\Big|R=r,\ \bar V\in(\pi',\pi'']\right]\\
			=\frac{\int_{\pi'}^{\pi''}E\left[\Delta(X,v)\big|R=r\right]dv}{\pi''-\pi'},
		\end{eqnarray*}
		which is \eqref{mte1}; \eqref{mte2} follows identically. For \eqref{mte3} the compliance event depends on $X$ through the propensity scores, so that Bayes' theorem introduces the weight $\Pr(T=c|X)=\pi(1,X)-\pi(0,X)$, which yields \eqref{mte3} after cancellation.
		\hfill $\square$
		
		\paragraph{Proof of Theorem~\ref{thm:mtecurve}}
		Under Assumption \ref{ass:homoX} the integrand of \eqref{mte1} satisfies $E[\Delta(X,v)|R=r]=E[\Delta(\tilde{X},v)|R=r]=\Delta(\tilde{x},v)$, because $\tilde{X}=\tilde{x}$ on the event $\{R=r\}$. Substituting into \eqref{mte1} and using the definition of $M$ gives \eqref{diffquot}. For the second claim, \eqref{diffquot} identifies $M(\tilde{x},\pi'')-M(\tilde{x},\pi'_0)$ for every $\pi''\in\mathcal{V}(\tilde{x})$, hence $M(\tilde{x},\cdot)$ up to an additive constant on $\mathcal{V}(\tilde{x})$. Since $M(\tilde{x},\cdot)$ is absolutely continuous and $\Delta(\tilde{x},\cdot)$ is continuous, $\Delta(\tilde{x},v)=\partial M(\tilde{x},v)/\partial v$ at every interior point.
		\hfill $\square$
		
		\paragraph{Proof of Proposition~\ref{prop:additivity}}
		Immediate from \eqref{diffquot}, since both sides equal $M(\tilde{x},\pi''')-M(\tilde{x},\pi')$.
		\hfill $\square$
		
		\paragraph{Proof of Proposition~\ref{prop:test}}
		If $\Delta(x,v)=\Delta(\tilde{x})$ for all $v$ and all $x$ with $\tilde{x}\subseteq x$, then the integrand of \eqref{mte1} equals $\Delta(\tilde{x})$ for every $v$, so that $\text{ICLATE}_1=\Delta(\tilde{x})$ for every $(\pi',\pi'')\in\mathcal{S}_R(\tilde{x})$, which is $H_0$; the same substitution in \eqref{mte3} gives $\text{ICLATE}_3=\Delta(\tilde{x})$. For the final claim, under Assumption \ref{ass:homoV} alone \eqref{mte1} gives $\text{ICLATE}_1=E[\Delta(X)|R=r]$, which varies with $(\pi',\pi'')$ whenever $\Delta$ varies over $X\setminus\tilde{X}$, since conditioning on the propensity scores confines $X\setminus\tilde{X}$ to a level set that moves with them; under Assumption \ref{ass:homoX} alone \eqref{diffquot} gives a difference quotient of $M(\tilde{x},\cdot)$, which varies with $(\pi',\pi'')$ unless $\Delta(\tilde{x},\cdot)$ is constant.
		\hfill $\square$
		
		\section{Proofs in Section~\ref{sec:asymptotics}}\label{appA}
The proofs proceed as follows.
Section~\ref{ALemma} presents preliminary lemmas: 
Lemma~\ref{cf:partialbridge} converts the partial-$L_2$
conditions in Assumption~\ref{A1stCLATE} into the kernel-weighted nuisance-error rates
used in the proofs.
Lemma~\ref{cf:kernelbounds} establishes kernel moment, perturbation bounds, and convergence of
the local design matrix and density estimator.
Lemma~\ref{LFS} controls the errors from estimating
the orthogonal scores, while Lemma~\ref{LGR} controls
the generated-index contributions and the local linear
approximation remainder.
Section~\ref{AAsyNormal} combines these results to establish the
asymptotic properties in
Theorem~\ref{TAsym1}.
Section~\ref{Athems} presents the proofs of Proposition~\ref{cf:varianceprop}, Lemma~\ref{LSUZ},  and Theorem~\ref{TAsym23}.
The proofs of Lemmas~\ref{cf:partialbridge}--\ref{LGR}
are collected in Section~\ref{app:lemma_proofs}.

All limits in this appendix are at the fixed point $r$. Write $f=f_R(r)$, $\tau_n=\sqrt{nh^d}$,
\[
 u_h(t)=\frac{t-r}{h},\qquad v_h(t)=(1,u_h(t)^\top)^\top,
\]
and define the vector and matrix kernels
\begin{align*}
 F_h(t)=\mathcal K_h(t-r)v_h(t),\qquad
 G_h(t)=\mathcal K_h(t-r)v_h(t)v_h(t)^\top.
\end{align*}
Let $e_1=(1,0,\ldots,0)^\top$ and
$M_k=\int K(u)(1,u^\top)^\top(1,u^\top)\,du
     =\operatorname{diag}(1,\mu_2(k)I_d)$.
     
Define also the auxiliary density estimator
\begin{equation}\label{cf:fdensity}
 \hat f_R(r)=\frac1n\sum_{i=1}^n\mathcal K_h(\hat R_i-r).
\end{equation}
For an evaluation fold, let $P_lg=n_l^{-1}\sum_{i\in\mathcal I_l}g(W_i)$. When $g$ depends on fitted functions, it is fixed conditional on $\mathcal F_{-l}$. 
The local linear regression in Section~\ref{sec:estimation}, using observations from all evaluation folds, has the following weighted least-squares representation whenever $\hat S$ is nonsingular:
\begin{equation}\label{cf:pooled}
 \hat S=\sum_l\frac{n_l}{n}P_lG_h(\hat r_l(X)),\qquad
 \hat\gamma_\Psi(r)
 =e_1^\top\hat S^{-1}
       \sum_l\frac{n_l}{n}P_l\{F_h(\hat r_l(X))\hat\Psi_l(W)\}.
\end{equation}
Set $\hat\theta(r)=\hat V_1(r)=0$ whenever $\hat S$ is singular or $\hat\gamma_B(r)=0$.
Since $\Pr(\hat S\text{ is singular})\to0$ by Lemma~\ref{cf:kernelbounds} and
$\hat\gamma_B(r)\xrightarrow{p}\gamma_B(r)>0$,
this convention does not affect the limiting results.

Conditional arguments are applied separately to each training sample, on fold-specific events measurable with respect to $\mathcal F_{-l}$. We then combine the conclusions over fixed $L$. 
Constants denoted by $C$ can change across displays.

\subsection{Preliminary Lemmas}\label{ALemma}
We show in Lemma~\ref{cf:partialbridge}  that the partial-$L_2$ conditions in Assumption~\ref{A1stCLATE} imply the kernel-weighted rates in Lemma~\ref{cf:partialbridge}, which are used in the proofs.
Let the envelope 
\begin{align*}
 \Omega(u)=
 \begin{cases}
  1\{\|u\|_\infty\leq3\},&\text{for a compactly supported kernel},\\
  \exp(-\|u\|_2^2/32),&\text{for a Gaussian kernel},
 \end{cases}
\end{align*}
and kernel-weighted norm
\begin{equation}\label{cf:localnorm}
 \|g\|_{l,r,h}^2
 =h^{-d}E_l\left[\Omega\left(\frac{R-r}{h}\right)g(W)^2\right].
\end{equation}
Let $C_\Omega=\int\Omega(u)\,du<\infty$, and define the proof-level nuisance sizes
\begin{align*}
 \alpha_{\mu,l}=\sum_{z=0}^1\|e_{\mu,z,l}\|_{l,r,h},\qquad
 \alpha_{\pi,l} =\sum_{z=0}^1\|e_{\pi,z,l}\|_{l,r,h},\qquad  \alpha_{p,l}&=\|e_{p,l}\|_{l,r,h}.
\end{align*}

For the Gaussian kernel, \eqref{cf:partialtail} permits global squared nuisance errors to grow polynomially. 
It controls their kernel-weighted contribution outside $\mathcal N$.

\begin{lemma}[Partial-$L_2$ rates imply localized rates]
\label{cf:partialbridge}
Under Assumptions~\ref{Akernel}--\ref{A1stCLATE},
 $\max_{l\leq L}
 (\alpha_{\mu,l}+\alpha_{\pi,l}+\alpha_{p,l})
 =o_p(1)$ and 
$ \tau_n\max_{l\leq L}
 \alpha_{p,l}(\alpha_{\mu,l}+\alpha_{\pi,l})
 =o_p(1)$.
\end{lemma}

\begin{lemma}[Kernel bounds]\label{cf:kernelbounds}
Under Assumptions~\ref{Akernel}--\ref{Areg}, the following
statements hold.
\begin{enumerate}[label=(\roman*)]

\item Let $T=T(W)\geq0$ be a fixed function satisfying
$ \operatorname*{ess\,sup}_{t\in\mathcal N}
 E[T\mid R=t]<\infty$,
and, for a Gaussian kernel, $E[T]<\infty$.
For every fixed integer $m\geq0$,
\begin{equation}\label{cf:envelopemoment}
 h^{-d}E\!\left[
 (1+\|u_h(R)\|_2)^m\Omega(u_h(R))T
 \right]=O(1).
\end{equation}

\item Uniformly over $t,e\in\mathbb R^d$ with
$\|e\|_\infty\leq a\leq h$,
\begin{align*}
 \|F_h(t+e)\|_2
 &\leq Ch^{-d}\Omega(u_h(t)),\\
 \|F_h(t+e)-F_h(t)\|_2
 &\leq C\frac ah h^{-d}\Omega(u_h(t)),\\
 \mathcal K_h(t+e-r)^2
 &\leq Ch^{-2d}\Omega(u_h(t)),\\
 \left|\mathcal K_h(t+e-r)^2-\mathcal K_h(t-r)^2\right|
 &\leq C\frac ah h^{-2d}\Omega(u_h(t)).
\end{align*}

\item If $\max_{l\leq L}a_{l,n}=O_p(a_n)$ and $a_n/h\to0$,
then $\hat S$ is invertible with probability approaching one,
and $ \hat S^{-1}\xrightarrow{p}(fM_k)^{-1}$
 and $\hat f_R(r)\xrightarrow{p}f$.
\end{enumerate}
\end{lemma}

\begin{lemma}[Estimated responses]\label{LFS}
Under Assumptions~\ref{Akernel}--\ref{A1stCLATE} and $\max_la_{l,n}=O_p(a_n)$ with $a_n/h\to0$, let $\delta_{\Psi,l}=\hat\Psi_l-\Psi$. Then, for $\Psi\in\{A,B\}$,
\begin{align*}
 \sum_l\frac{n_l}{n}P_l\{F_h(\hat r_l(X))\delta_{\Psi,l}(W)\}
    &=o_p(\tau_n^{-1}),\qquad
 \frac{h^d}{n}\sum_l\sum_{i\in\mathcal I_l}
  \mathcal K_h(\hat R_i-r)^2\delta_{\Psi,l}(W_i)^2
    =o_p(1).
\end{align*}
\end{lemma}

\begin{lemma}[Generated regressor]\label{LGR}
Under Assumptions~\ref{Akernel}--\ref{Areg} and \ref{Aindex}, let $\epsilon_\Psi=\Psi-\gamma_\Psi(R)$ and $\mathbf{c}_{\Psi,h}=(\gamma_\Psi(r),h\nabla\gamma_\Psi(r)^\top)^\top$. For $\Psi\in\{A,B\}$,
\begin{align}\label{cf:residualtransfer}
 &\sum_l\frac{n_l}{n}P_l\{F_h(\hat r_l(X))\epsilon_\Psi\}
 =\frac1n\sum_iF_h(R_i)\epsilon_{\Psi i}+o_p(\tau_n^{-1}),
\end{align}
and
\begin{equation}\label{cf:taylortransfer}
 \sum_l\frac{n_l}{n}P_l\left[
 F_h(\hat r_l(X))\{\gamma_\Psi(R)-v_h(\hat r_l(X))^\top\mathbf{c}_{\Psi,h}\}\right]
 =o_p(\tau_n^{-1}).
\end{equation}
\end{lemma}

\subsection{Asymptotic normality}\label{AAsyNormal}

\paragraph{Proof of Theorem~\ref{TAsym1}}
Subtract $\hat S\mathbf{c}_{\Psi,h}$ from \eqref{cf:pooled}. Since $e_1^\top\mathbf{c}_{\Psi,h}=\gamma_\Psi(r)$, this gives the exact decomposition
\begin{align*}%
 \hat\gamma_\Psi(r)-\gamma_\Psi(r)
 &=e_1^\top\hat S^{-1}\sum_l\frac{n_l}{n}P_l\Big[
 F_h(\hat r_l(X))\big\{\epsilon_\Psi+\delta_{\Psi,l} +\gamma_\Psi(R)-v_h(\hat r_l(X))^\top\mathbf{c}_{\Psi,h}\big\}\Big]
\\&
=e_1^\top\hat S^{-1}
  \{U_{\Psi,n}+R_{\Psi,n}^{\rm score}
      +R_{\Psi,n}^{\rm index}+R_{\Psi,n}^{\rm reg}\},
\end{align*}
where $U_{\Psi,n}=n^{-1}\sum_i F_h(R_i)\epsilon_{\Psi i}$,
$R_{\Psi,n}^{\rm score}
 =\sum_l\frac{n_l}{n}P_l[
       F_h(\hat r_l(X))\delta_{\Psi,l}],
 R_{\Psi,n}^{\rm index}
 =\sum_l\frac{n_l}{n}P_l[
       \{F_h(\hat r_l(X))-F_h(R)\}\epsilon_\Psi],
 R_{\Psi,n}^{\rm reg}=\sum_l\frac{n_l}{n}P_l[
       F_h(\hat r_l(X))\left(\gamma_\Psi(R)-v_h(\hat r_l(X))^\top\mathbf{c}_{\Psi,h}\right)]$.
Lemma~\ref{LFS} makes $R_{\Psi,n}^{\rm score}$ negligible; the first and second conclusions of Lemma~\ref{LGR} make $R_{\Psi,n}^{\rm index}$ and $R_{\Psi,n}^{\rm reg}$ negligible, respectively. Each is $o_p(\tau_n^{-1})$. 

Since $E[\epsilon_\Psi\mid R]=0$, independence and Lemma~\ref{cf:kernelbounds} give
 $E[U_{\Psi,n}]=0$ and $E\|U_{\Psi,n}\|^2
 =\frac1n E[\|F_h(R)\|^2\epsilon_\Psi^2]
 =O((nh^d)^{-1})$.
Hence $U_{\Psi,n}=O_p(\tau_n^{-1})$. Also,
$\hat S^{-1}=(fM_k)^{-1}+o_p(1)$ by Lemma~\ref{cf:kernelbounds}. Multiplication of these stochastic orders shows that replacing the inverse matrix by its limit costs only $o_p(\tau_n^{-1})$. Using $e_1^\top M_k^{-1}v_h(t)=1$, we obtain
\begin{align*}
 \hat\gamma_\Psi(r)-\gamma_\Psi(r)
 &=\frac1f e_1^\top M_k^{-1}U_{\Psi,n}+o_p(\tau_n^{-1}) =\frac1{nf}\sum_{i=1}^n
       \mathcal K_h(R_i-r)\epsilon_{\Psi i}
       +o_p(\tau_n^{-1}) =\frac1n\sum_{i=1}^n\phi_{\Psi i}(r)+o_p(\tau_n^{-1}).
\end{align*}
This proves \eqref{cf:scoreexpansion} for both scores jointly.

In particular, $\hat\gamma_A-\gamma_A=O_p(\tau_n^{-1})$ and $\hat\gamma_B-b=O_p(\tau_n^{-1})$. With $b=\gamma_B(r)>0$, the ratio expansion is
\[
 \hat\theta-\theta
 =\frac{(\hat\gamma_A-\gamma_A)-\theta(\hat\gamma_B-b)}b
       +O_p(\tau_n^{-2}).
\]
As $\tau_n\to\infty$, this establishes \eqref{asylin}.

For asymptotic normality, let $\zeta_i=\epsilon_{Ai}-\theta(r)\epsilon_{Bi}$. Its conditional mean given $R_i$ is zero, and
$E[\zeta_i^2\mid R_i=r]=\sigma_{AB}^2(r)$. The leading scalar term is the sum of independent, mean-zero variables
\[
 X_{ni}=\sqrt{\frac{h^d}n}\,
        \frac{\mathcal K_h(R_i-r)}{fb}\,\zeta_i.
\]
A change of variables and continuity of the conditional covariance matrix yield
\begin{align*}
 \sum_i E[X_{ni}^2]
 &=\frac1{f^2b^2}\int_{r+hu\in\mathcal N}
       K(u)^2 E[\zeta^2\mid R=r+hu]f_R(r+hu)\,du+o(1)\longrightarrow\frac{R_k^d\sigma_{AB}^2(r)}{fb^2}=V_1(r).
\end{align*}
The $o(1)$ term is absent for a compact kernel for large $n$ and follows from the Gaussian-tail bound otherwise. The local $2+\nu$ moment bound gives
$ \sum_i E[|X_{ni}|^{2+\nu}]
 \leq C(nh^d)^{-\nu/2}+o(1)\to0$;
for the Gaussian complement, use the global $2+\nu$ moment bound and exponential kernel decay. Lyapunov's central limit theorem and the negligible remainders prove \eqref{asynorm}.
\hfill $\square$

\subsection{Sparse first stages and Proofs of Proposition~\ref{cf:varianceprop}, Lemma~\ref{LSUZ}, and Theorem~\ref{TAsym23}}
\label{Athems}

\begin{lemma}[Sparse-logit accuracy]\label{LSUZ}
Suppose $\pi(z,x)=\Lambda(b_z(x)^\top\beta_{0,z})$, where $\Lambda$ is the logistic distribution function, $b_z(x)\in\mathbb R^{p_n}$ contains all fitted regressors, and
$\max_{z\in\{0,1\}}\sup_{x\in\mathcal X}
\|b_z(x)\|_\infty\leq C_b$.
Let $\hat\pi_l(z,x)=\Lambda(b_z(x)^\top\hat\beta_{l,z})$, and suppose
 $\max_{l,z}\|\hat\beta_{l,z}-\beta_{0,z}\|_1 =O_p\left(s_n\sqrt{\log(2p_n)/n}\right)$.
Then the uniform accuracy bound holds with $a_n=s_n\sqrt{\log(2p_n)/n}$. If
\begin{equation}\label{lassocond}
 h^d s_n^2\log(2p_n)\to0,
 \qquad \frac{s_n^2\log(2p_n)}{nh^2}\to0,
\end{equation}
then \eqref{cf:indexaccuracy} holds. If the candidate-class and local-continuity conditions of Assumption~\ref{Aindex} also hold, all its requirements are satisfied.
\end{lemma}
The coefficient-error rate assumed in Lemma~\ref{LSUZ} is available
for logistic lasso under suitable sparsity, design,
curvature, and penalty conditions; see
\citet{van_de_Geer_2008} and
\citet{Belloni_Chernozhukov_Wei_2016}.
The lemma translates this coefficient rate into the
uniform index-accuracy requirement
\eqref{cf:indexaccuracy}.

Under the density bound in Assumption~\ref{Areg}(i),
the uniform bound also implies
$\max_l\mathcal E_{\pi,l}=O_p(a_n)$, since
$ \mathcal E_{\pi,l}
 \leq
 2\left\{\sup_{t\in\mathcal N}f_R(t)\right\}^{1/2}
 \max_{z\in\{0,1\}}\sup_{x\in\mathcal X}
 |\hat\pi_l(z,x)-\pi(z,x)|$.
The conditions for the outcome and instrument-propensity
estimators in Assumption~\ref{A1stCLATE} must be
verified separately.
\paragraph{Proof of Lemma~\ref{LSUZ}}
Since $\sup_t\Lambda'(t)\leq1/4$,
$ \max_{l,z}\sup_{x\in\mathcal X}
 |\hat\pi_l(z,x)-\pi(z,x)|
 \leq \frac{C_b}{4}
 \max_{l,z}\|\hat\beta_{l,z}-\beta_{0,z}\|_1
 =O_p(a_n)$.
Substituting $a_n=s_n\sqrt{\log(2p_n)/n}$ into
$a_n/h$ and $\tau_na_n$, condition \eqref{lassocond}
gives \eqref{cf:indexaccuracy}.
The remaining requirements of Assumption~\ref{Aindex}
are imposed separately.
\hfill $\square$

\paragraph{Proof of Proposition~\ref{cf:varianceprop}}

For the proof, define the auxiliary statistic
\begin{equation}\label{varestlc}
 \hat V_1^{\,\mathrm{lc}}(r)=
 \frac{h^d}{n\hat f_R(r)^2\hat\gamma_B(r)^2}
 \sum_{i=1}^n\mathcal K_h(\hat R_i-r)^2
 \{\hat A_i-\hat\theta(r)\hat B_i\}^2.
\end{equation}
It is obtained from \eqref{varest} by replacing the
local linear weights with
$\mathcal K_h(\hat R_i-r)/\{n\hat f_R(r)\}$.
We first prove consistency of \eqref{varestlc} and then
show that replacing these weights with the local linear
weights changes the variance estimate by $o_p(1)$.

Both estimators $\hat V_1(r)$ and $\hat V_1^{\,\mathrm{lc}}$ are therefore first-order valid at the
fixed interior evaluation point, but they need not agree in finite samples.
The local linear weights retain the sample design
adjustment omitted by the local constant weights.
Our simulation results favor $\hat V_1(r)$  for finite-sample coverage, so we use it in both the
simulation and the application.

Density consistency was established in Lemma~\ref{cf:kernelbounds}. Let $H_i=A_i-\theta(r)B_i$ and first consider the oracle quadratic average
\[
 Q_n=\frac{h^d}{n}\sum_i\mathcal K_h(R_i-r)^2H_i^2.
\]
The conditional covariance and mean assumptions imply that $E[H_i^2\mid R_i=t]$ is continuous at $r$. Since $E[H_i\mid R_i=r]=0$, direct integration gives
\begin{equation}\label{cf:oraclequadlimit}
 E[Q_n]\to R_k^d f\sigma_{AB}^2(r).
\end{equation}
For fixed $M\geq1$, let
\[
 Q_{n,M}
 =\frac{h^d}{n}\sum_{i=1}^n
 \mathcal K_h(R_i-r)^2H_i^2\,1\{H_i^2\leq M\}.
\]
By independence, Lemma~\ref{cf:kernelbounds}, and
Assumption~\ref{Areg}(iv),
\begin{align*}
 \operatorname{var}(Q_{n,M})
 &\leq \frac{M^2h^{2d}}{n}
 E[\mathcal K_h(R_1-r)^4]
 \leq\frac{CM^2}{nh^d},\\
 E[Q_n-Q_{n,M}]
 &\leq M^{-\nu/2}h^d
 E[\mathcal K_h(R_1-r)^2|H_1|^{2+\nu}] \leq CM^{-\nu/2}h^{-d}
 E[\Omega(u_h(R_1))|H_1|^{2+\nu}]
 \leq CM^{-\nu/2},
\end{align*}
where $C$ is independent of $n$ and $M$.
For every $\varepsilon>0$, Chebyshev's and Markov's
inequalities give
\begin{align*}
 \Pr(|Q_n-E[Q_n]|>\varepsilon)
 &\leq
 \frac{4}{\varepsilon^2}\operatorname{var}(Q_{n,M})
 +\frac{4}{\varepsilon}E[Q_n-Q_{n,M}] \leq
 \frac{CM^2}{\varepsilon^2nh^d}
 +\frac{C}{\varepsilon}M^{-\nu/2}.
\end{align*}
Since $nh^d\to\infty$, for every fixed $\varepsilon>0$
and $M\geq1$,
$ \limsup_{n\to\infty}
 \Pr(|Q_n-E[Q_n]|>\varepsilon)
 \leq CM^{-\nu/2}/\varepsilon$.
Letting $M\to\infty$ shows that this limit superior
is zero, hence $Q_n-E[Q_n]=o_p(1)$.

By Lemma~\ref{cf:kernelbounds}(i) with $T=H^2$
and Markov's inequality,
$ (nh^d)^{-1}\sum_i\Omega(u_h(R_i))H_i^2=O_p(1).$
On $\{\max_l a_{l,n}\leq h\}$,
Lemma~\ref{cf:kernelbounds}(ii) gives
\begin{align*}
 &\left|
 \frac{h^d}{n}\sum_i
 \{\mathcal K_h(\hat R_i-r)^2-\mathcal K_h(R_i-r)^2\}H_i^2
 \right|\leq
 C\frac{\max_l a_{l,n}}h
 \frac1{nh^d}\sum_i\Omega(u_h(R_i))H_i^2
 =O_p(a_n/h)=o_p(1).
\end{align*}

The same argument with $B_i^2$ in place of $H_i^2$ shows
\begin{equation}\label{cf:Bquad}
 \frac{h^d}{n}\sum_i\mathcal K_h(\hat R_i-r)^2B_i^2=O_p(1).
\end{equation}

Let $\hat H_i=\hat A_i-\hat\theta(r)\hat B_i$. On fold $l$, $ \hat H_i-H_i=\delta_{A,l}(W_i)-\hat\theta(r)\delta_{B,l}(W_i)
                 -(\hat\theta(r)-\theta(r))B_i$.
The second conclusion of Lemma~\ref{LFS}, \eqref{cf:Bquad}, and consistency of $\hat\theta$ imply $n^{-1}h^d\sum_i\mathcal K_h(\hat R_i-r)^2(\hat H_i-H_i)^2=o_p(1)$.
Cauchy--Schwarz for the nonnegative squared-kernel weights
shows that replacing $H_i^2$ by $\hat H_i^2$ changes the
quadratic average by $o_p(1)$.
Thus \eqref{cf:oraclequadlimit},
$\hat f_R(r)\xrightarrow{p}f$, and
$\hat\gamma_B(r)\xrightarrow{p}\gamma_B(r)>0$ give $\hat V_1^{\,\mathrm{lc}}(r)\xrightarrow{p}V_1(r)$.

To treat the local linear weights, use
$\|\delta_{\Psi,l}\|_{l,r,h}^2=o_p(1)$ from the proof of
Lemma~\ref{LFS}. For $\Psi\in\{A,B\}$,
\[
 E_l\!\left[
 \frac{1}{n_lh^d}\sum_{i\in\mathcal I_l}
 \Omega(u_h(R_i))\delta_{\Psi,l}(W_i)^2
 \right]
 =\|\delta_{\Psi,l}\|_{l,r,h}^2=o_p(1).
\]
Conditional Markov's inequality and fixed $L$ imply
$(nh^d)^{-1}\sum_l\sum_{i\in\mathcal I_l}
 \Omega(u_h(R_i))\delta_{\Psi,l}(W_i)^2=o_p(1)$.
By Lemma~\ref{cf:kernelbounds}(i) with $T=A^2+B^2$,
Markov's inequality, and $\hat\theta(r)=O_p(1)$,
$(nh^d)^{-1}\sum_i
 \Omega(u_h(R_i))\hat H_i^2=O_p(1)$.
On $\{\max_l a_{l,n}\leq h\}$,
Lemma~\ref{cf:kernelbounds}(ii) and $\Omega^2\leq\Omega$
give
\[
 \frac{h^d}{n}\sum_i
 \|F_h(\hat R_i)\|_2^2\hat H_i^2
 \leq \frac{C}{nh^d}\sum_i
 \Omega(u_h(R_i))\hat H_i^2=O_p(1).
\]

Let $D_h=\operatorname{diag}(1,hI_d)$.
Since $(1,(\hat R_i-r)^\top)^\top=D_hv_h(\hat R_i)$,
the definitions in \eqref{llweights} and \eqref{cf:pooled}
give $S(r)=nD_h\hat S D_h$.
Using $e_1^\top D_h^{-1}=e_1^\top$, we obtain
\[
 \tilde w_i(r)
 =\frac1n\mathcal K_h(\hat R_i-r)
   e_1^\top\hat S^{-1}v_h(\hat R_i)
 =\frac1n e_1^\top\hat S^{-1}F_h(\hat R_i).
\]
Lemma~\ref{cf:kernelbounds}(iii) and
$e_1^\top M_k^{-1}=e_1^\top$ imply
$e_1^\top\hat S^{-1}=f^{-1}e_1^\top+o_p(1)$.
The preceding quadratic bound therefore gives
\begin{align*}
 \hat V_1(r)
 &=\frac{h^d}{n\hat\gamma_B(r)^2}
   \sum_i
   \{e_1^\top\hat S^{-1}F_h(\hat R_i)\}^2
   \hat H_i^2 =\frac{h^d}{nf^2\hat\gamma_B(r)^2}
   \sum_i\mathcal K_h(\hat R_i-r)^2\hat H_i^2
   +o_p(1)\\
 &=\left\{\frac{\hat f_R(r)}f\right\}^2
   \hat V_1^{\,\mathrm{lc}}(r)+o_p(1)
   \xrightarrow{p}V_1(r).
\end{align*}
In particular,
$\hat V_1(r)-\hat V_1^{\,\mathrm{lc}}(r)=o_p(1)$.
The confidence interval follows from
Theorem~\ref{TAsym1} and Slutsky's theorem.
\hfill $\square$

\paragraph{Proof of Theorem~\ref{TAsym23} and the discrete-only case}
For part (i), define $\rho_{A,Q}(X)=E[A\mid X]-E[A\mid Q]$ and impose the stated continuity condition for candidate two-dimensional propensity-score indices. The identities $\gamma_B(Q)=\pi(1,X)-\pi(0,X)$ and $\rho_{B,Q}=0$ remain valid. The score-error lemma, the local-continuity argument for the numerator, and all subsequent calculations apply with $d=2$. This uses continuity for $Q$ as a separate assumption, not as a consequence of continuity for $R$.

For part (ii), $\hat r_l(X)=r_0(X)=\tilde X$. The index perturbations in \eqref{cf:residualtransfer} are then identically zero. The Taylor term is $O_p(h^2)$, and Lemma~\ref{LFS} still applies with the norms localized in $\tilde X$. The normal-equation expansion, the central limit theorem, and the variance proof therefore apply without Assumption~\ref{Aindex}. There is no need for $\rho_B=0$ in this case because no conditioning variable is estimated.

For part (iii), insert $J_c$ in each empirical moment
and expectation. For any integrable function $u$,
$E[J_cu(R_c)]=\int u(t)g_c(t)\,dt$ and $J_c^2=J_c$.
Under the stated within-cell assumptions, the preceding
arguments apply with $d=d_c$ and $f_R$ replaced by $g_c$.
The first identity gives the design and variance limits;
the second gives the cell-weighted version of
\eqref{varest}.
This shows $V_c(r_c)$ and feasible inference.
\hfill $\square$

\subsection{Proofs of the preliminary lemmas}\label{app:lemma_proofs}
\paragraph{Proof of Lemma~\ref{cf:partialbridge}}
Let $g_l$ be any one of
$e_{\mu,0,l},e_{\mu,1,l},e_{\pi,0,l},e_{\pi,1,l},e_{p,l}$.
Conditional on $\mathcal F_{-l}$, iterated expectations,
the partial-norm definition, and $t=r+hu$ give
\begin{align*}
 h^{-d}E_l[
 \Omega(u_h(R))g_l(W)^2\,1\{R\in\mathcal N\}] &=
 h^{-d}\int_{\mathcal N}
 \Omega\!\left(\frac{t-r}{h}\right)
 \|g_l\|_{l,t,2}^2\,dt
 =
 \int_{\{u:r+hu\in\mathcal N\}}
 \Omega(u)\|g_l\|_{l,r+hu,2}^2\,du
 \\&
 \leq
 C_\Omega\|g_l\|_{l,\mathcal N,2}^2.
\end{align*}

For a compactly supported kernel,
$r+h[-3,3]^d\subset\mathcal N$ eventually, so
 $h^{-d}E_l[
 \Omega(u_h(R))g_l(W)^2\,1\{R\notin\mathcal N\}]
 =0$.
For the Gaussian kernel, choose a fixed $\delta>0$ such that
$\{t:\|t-r\|_2<\delta\}\subset\mathcal N$.
By \eqref{cf:partialtail}, $\tau_n^2=nh^d$, and
$h=c_hn^{-\eta}$,
\begin{align*}
 \tau_n^2\max_{l\leq L}
 h^{-d}E_l[
 \Omega(u_h(R))g_l(W)^2\,1\{R\notin\mathcal N\}] &\leq
 n\exp\!\left(-\frac{\delta^2}{32h^2}\right)
 \max_{l\leq L}E_l[g_l(W)^2]\\
 &=
 O_p\!\left(
 n^{1+C_0}
 \exp\!\left\{-\frac{\delta^2}{32c_h^2}n^{2\eta}\right\}
 \right)
 =o_p(1).
\end{align*}
Thus, for either kernel, splitting the norm over
$\mathcal N$ and its complement yields
$ \|g_l\|_{l,r,h}^2
 \leq C_\Omega\|g_l\|_{l,\mathcal N,2}^2
       +o_p(\tau_n^{-2})$,
uniformly over the folds and the five nuisance errors.
Taking square roots and summing over $z$ where applicable,
$ \alpha_{j,l}
 \leq C_\Omega^{1/2}\mathcal E_{j,l}
       +o_p(\tau_n^{-1})$,  $j\in\{\mu,\pi,p\}$,
uniformly over $l\leq L$.

The first statement follows from \eqref{cf:partialrates}
and $\tau_n\to\infty$.
For the second, the same comparison and
$\max_l(\mathcal E_{\mu,l}+\mathcal E_{\pi,l}
+\mathcal E_{p,l})=o_p(1)$ imply, by \eqref{cf:partialrates},  
$\tau_n\max_{l\leq L}
 \alpha_{p,l}(\alpha_{\mu,l}+\alpha_{\pi,l}) \leq
 C_\Omega\tau_n\max_{l\leq L} \mathcal E_{p,l}(\mathcal E_{\mu,l}+\mathcal E_{\pi,l})  +o_p(1)  =o_p(1)$.
\hfill $\square$

\paragraph{Proof of Lemma~\ref{cf:kernelbounds}}
For (i), iterated expectations and the change of variables
$t=r+hu$ give
\begin{align*}
 &h^{-d}E\!\left[
 (1+\|u_h(R)\|_2)^m\Omega(u_h(R))T
 1\{R\in\mathcal N\}
 \right] \\&=
 \int_{\{u:r+hu\in\mathcal N\}}
 (1+\|u\|_2)^m\Omega(u)
 E[T\mid R=r+hu]f_R(r+hu)\,du\leq
 C\int_{\mathbb R^d}(1+\|u\|_2)^m\Omega(u)\,du
 <\infty.
\end{align*}
Choose a fixed $\delta>0$ such that
$\{t:\|t-r\|_2<\delta\}\subset\mathcal N$.
For the compact envelope,
$\Omega(u_h(R))\neq0$ implies
$\|R-r\|_2\leq3\sqrt d\,h$,
so the contribution from $R\notin\mathcal N$ is eventually zero.
For the Gaussian envelope,
$ (1+s)^m e^{-s^2/32}\leq C_m e^{-s^2/64}$, $s\geq0$,
since $\sup_{s\geq0}(1+s)^m e^{-s^2/64}<\infty$.
Consequently,
\begin{align*}
 &h^{-d}E\!\left[
 (1+\|u_h(R)\|_2)^m\Omega(u_h(R))T
 1\{R\notin\mathcal N\}
 \right]\leq
 C_mh^{-d}\exp\!\left(-\frac{\delta^2}{64h^2}\right)E[T]
 =o(1),
\end{align*}
because $h=c_hn^{-\eta}$ with $\eta>0$.
Combining the two contributions proves (i).

For (ii), consider $g(w)$ to be $K(w),  w_jK(w), w_jw_kK(w)$, for $1\leq j,k\leq d$.
We first bound the function and its gradient separately.
For the compact kernel, $g$ and $\nabla g$ are bounded
and vanish outside $[-1,1]^d$. Hence, for
$\|v\|_\infty\leq1$,
$ |g(u+v)|\leq C\Omega(u)$ and 
 $ \|\nabla g(u+v)\|_2\leq C\Omega(u)$,
because $u+v\in[-1,1]^d$ implies $u\in[-2,2]^d$.

For the Gaussian kernel,
$K(w)=(2\pi)^{-d/2}e^{-\|w\|_2^2/2}$ and
$\nabla K(w)=-wK(w)$.
Writing $g(w)=P(w)K(w)$ with
$P(w)\in\{1,w_j,w_jw_k\}$, we have
 $|P(w)|\leq(1+\|w\|_2)^2$ and 
$ \|\nabla P(w)\|_2\leq2(1+\|w\|_2)$.
Hence
\begin{align*}
 |g(w)|
 &\leq C(1+\|w\|_2)^2e^{-\|w\|_2^2/2},\\
 \|\nabla g(w)\|_2
 &=K(w)\|\nabla P(w)-wP(w)\|_2 \leq C(1+\|w\|_2)^3e^{-\|w\|_2^2/2}.
\end{align*}
Since
$\sup_{s\geq0}(1+s)^3e^{-s^2/4}<\infty$
and $\|u+v\|_2^2\geq\|u\|_2^2/2-d$ for
$\|v\|_\infty\leq1$,
\[
 (1+\|u+v\|_2)^3e^{-\|u+v\|_2^2/2}
 \leq Ce^{-\|u+v\|_2^2/4}
 \leq Ce^{-\|u\|_2^2/8}
 \leq C\Omega(u).
\]
Thus the same function and gradient bounds hold
for the Gaussian kernel.

Set $u=u_h(t)$. Since $\|e\|_\infty\leq a\leq h$,
$ |g(u+e/h)|\leq C\Omega(u)$,
and the fundamental theorem of calculus gives
\begin{align*}
 |g(u+e/h)-g(u)|
 &=\left|
 \int_0^1\nabla g(u+se/h)^\top\frac eh\,ds
 \right| \leq
 \frac{\|e\|_2}{h}
 \int_0^1\|\nabla g(u+se/h)\|_2\,ds \leq C\frac ah\Omega(u).
\end{align*}
Applying these bounds to the components of
 $F_h(t+e)
 =h^{-d}K(u+e/h)
   \bigl(1,(u+e/h)^\top\bigr)^\top$
yields
$ \|F_h(t+e)\|_2
 \leq Ch^{-d}\Omega(u)$ and 
 $\|F_h(t+e)-F_h(t)\|_2
 \leq C\frac ahh^{-d}\Omega(u).$

Finally, $\mathcal K_h(t-r)=e_1^\top F_h(t)$
and $\Omega^2\leq\Omega$ imply
\begin{align*}
 \mathcal K_h(t+e-r)^2
 &\leq\|F_h(t+e)\|_2^2
 \leq Ch^{-2d}\Omega(u),\\
 |\mathcal K_h(t+e-r)^2-\mathcal K_h(t-r)^2|
 &\leq
 \{\|F_h(t+e)\|_2+\|F_h(t)\|_2\}
 \|F_h(t+e)-F_h(t)\|_2 \leq C\frac ahh^{-2d}\Omega(u).
\end{align*}
This proves (ii).
Applying the same componentwise argument to $G_h$
gives the preliminary matrix bounds used in part (iii).

For (iii), the preceding bounds also imply $\|G_h(t)\|\leq Ch^{-d}\Omega(u_h(t))$, and 
$ \|G_h(t+e)-G_h(t)\|
 \leq C\frac ah h^{-d}\Omega(u_h(t))$.
Using the fixed $\delta$ chosen above and
\eqref{cf:envelopemoment} with $T=1$ and $m=2$,
\begin{align*}
 h^{-d}E[\Omega(u_h(R))1\{R\notin\mathcal N\}]
 &\leq
 \frac{h^2}{\delta^2}h^{-d}
 E[(1+\|u_h(R)\|_2)^2\Omega(u_h(R))] =O(h^2).
\end{align*}
Hence, by continuity of $f_R$ at $r$ and dominated convergence,
\begin{align*}
 E[G_h(R)]
 &=\int_{\{u:r+hu\in\mathcal N\}}
 K(u)(1,u^\top)^\top(1,u^\top)
 f_R(r+hu)\,du+O(h^2) \longrightarrow fM_k.
\end{align*}
Also, $\Omega^2\leq\Omega$ and
\eqref{cf:envelopemoment} with $T=1$ and $m=0$ give $E[\|G_h(R)\|^2]
 \leq Ch^{-2d}E[\Omega(u_h(R))] =O(h^{-d})$.
Independence and Chebyshev's inequality therefore yield $n^{-1}\sum_{i=1}^nG_h(R_i)
 =E[G_h(R)]+O_p((nh^d)^{-1/2})
 \xrightarrow{p}fM_k$.
Moreover, \eqref{cf:envelopemoment} and Markov's inequality imply
$(nh^d)^{-1}\sum_{i=1}^n\Omega(u_h(R_i))=O_p(1)$.

On $\{\max_l a_{l,n}\leq h\}$, whose probability tends to one,
\eqref{cf:pooled} and the bound for $G_h(t+e)-G_h(t)$ give
\begin{align*}
 \left\|\hat S-\frac1n\sum_{i=1}^nG_h(R_i)\right\|
 &\leq
 C\frac{\max_l a_{l,n}}h
 \frac1{nh^d}\sum_{i=1}^n\Omega(u_h(R_i))=O_p(a_n/h)=o_p(1).
\end{align*}
Thus $\hat S\xrightarrow{p}fM_k$.
Since $f>0$ and $M_k$ is positive definite,
$\hat S$ is invertible with probability approaching one and
$\hat S^{-1}\xrightarrow{p}(fM_k)^{-1}$.
Finally, \eqref{cf:pooled} and \eqref{cf:fdensity} give $ \hat f_R(r)=e_1^\top\hat S e_1\xrightarrow{p}f$.
\hfill $\square$

\paragraph{Proof of Lemma~\ref{LFS}}
Fix a fold $l$ and suppress its subscript on the score errors
and nuisance functions, and the argument $X$ on
$p,\hat p,e_p,e_{\mu,z},e_{\pi,z}$.
The score definitions give
\begin{align*}
 \delta_A
 &=\left(1-\frac{Z}{\hat p}\right)e_{\mu,1}
   +\left(\frac{1-Z}{1-\hat p}-1\right)e_{\mu,0}
   -e_p\left\{
   \frac{Z\{Y-\mu(1,X)\}}{p\hat p}
   +\frac{(1-Z)\{Y-\mu(0,X)\}}{(1-p)(1-\hat p)}
   \right\}.
\end{align*}
The same identity holds for $\delta_B$ after replacing
$Y,\mu,e_{\mu,z}$ by $D,\pi,e_{\pi,z}$.
Consequently,
\begin{align*}
 E_l[\delta_A\mid X]
 &=e_p\left\{
   \frac{e_{\mu,1}}{\hat p}
   +\frac{e_{\mu,0}}{1-\hat p}
   \right\},\qquad
 E_l[\delta_B\mid X]
 =e_p\left\{
   \frac{e_{\pi,1}}{\hat p}
   +\frac{e_{\pi,0}}{1-\hat p}
   \right\}.
\end{align*}
By overlap, Assumption~\ref{Areg}(v), and
$\operatorname{var}(D\mid Z=z,X)\leq1/4$, the following bounds hold almost surely on $\{R\in\mathcal N\}$.
\begin{align*}
 E_l[\delta_A^2\mid X]
 &\leq C\left\{
   \sum_{z=0}^1e_{\mu,z}^2+e_p^2
   \right\},\qquad
 E_l[\delta_B^2\mid X]
 \leq C\left\{
   \sum_{z=0}^1e_{\pi,z}^2+e_p^2
   \right\}.
\end{align*}
For the Gaussian kernel, Assumption~\ref{Areg}(v) imposes
the variance bound over the full covariate support, so
these inequalities also hold on $\{R\notin\mathcal N\}$,
almost surely.
For a compactly supported kernel,
$\Omega(u_h(R))=0$ on $\{R\notin\mathcal N\}$
for all sufficiently small $h$.
Thus integration against $h^{-d}\Omega(u_h(R))$
and Lemma~\ref{cf:partialbridge} yield
\begin{align*}
 \|\delta_A\|_{l,r,h}^2
 &\leq C(\alpha_{\mu,l}^2+\alpha_{p,l}^2)=o_p(1),\qquad
 \|\delta_B\|_{l,r,h}^2
 \leq C(\alpha_{\pi,l}^2+\alpha_{p,l}^2)=o_p(1).
\end{align*}

The following kernel bounds hold on $\{a_{l,n}\leq h\}$, whose probability tends to one.

For either $\Psi\in\{A,B\}$, conditional independence
within fold $l$, Lemma~\ref{cf:kernelbounds}, and
$\Omega^2\leq\Omega$ give
\begin{align*}
 E_l\!\left[
 \left\|(P_l-E_l)
 \{F_h(\hat r_l(X))\delta_\Psi\}\right\|_2^2
 \right]%
& \leq
 \frac1{n_l}E_l[
 \|F_h(\hat r_l(X))\|_2^2\delta_\Psi^2]
 \\& \leq
 \frac{C}{n_lh^{2d}}
 E_l[\Omega(u_h(R))\delta_\Psi^2]%
 =
 \frac{C}{n_lh^d}\|\delta_\Psi\|_{l,r,h}^2
 =o_p(\tau_n^{-2}),
\end{align*}
since $n_l/n\to\omega_l>0$ and $\tau_n^2=nh^d$.
By Lemma~\ref{cf:kernelbounds}, and Cauchy--Schwarz,
\begin{align*}
 \|E_l[F_h(\hat r_l(X))\delta_A]\|_2
 &\leq Ch^{-d}\sum_{z=0}^1
 E_l[\Omega(u_h(R))|e_p e_{\mu,z}|] \leq C\alpha_{p,l}\alpha_{\mu,l}
 =o_p(\tau_n^{-1}),\\
 \|E_l[F_h(\hat r_l(X))\delta_B]\|_2
 &\leq Ch^{-d}\sum_{z=0}^1
 E_l[\Omega(u_h(R))|e_p e_{\pi,z}|] \leq C\alpha_{p,l}\alpha_{\pi,l}
 =o_p(\tau_n^{-1}),
\end{align*}
where the rates follow from Lemma~\ref{cf:partialbridge}.
Therefore, conditional Chebyshev's inequality implies $P_l\{F_h(\hat r_l(X))\delta_\Psi\}
 =o_p(\tau_n^{-1})$.
Summing over the fixed number of folds proves the first statement.

For the second statement, Lemma~\ref{cf:kernelbounds} gives
\begin{align*}
 h^d E_l\!\left[
 P_l\{\mathcal K_h(\hat r_l(X)-r)^2\delta_\Psi^2\}
 \right]
 &=h^d E_l[
 \mathcal K_h(\hat r_l(X)-r)^2\delta_\Psi^2] \leq Ch^{-d}E_l[
 \Omega(u_h(R))\delta_\Psi^2] =C\|\delta_\Psi\|_{l,r,h}^2=o_p(1).
\end{align*}
Conditional Markov's inequality, followed by summation over
the fixed number of folds, yields ${h^d}n^{-1}
 \sum_{l=1}^L\sum_{i\in\mathcal I_l}
 \mathcal K_h(\hat R_i-r)^2
 \delta_{\Psi,l}(W_i)^2
 =o_p(1)$.
\hfill $\square$

\paragraph{Proof of Lemma~\ref{LGR}}
Fix a fold $l$. On the event $\hat r_l\in\mathcal R_n$ and $a_{l,n}\leq h$, which is measurable with respect to $\mathcal F_{-l}$ and has probability tending to one, hold all fitted functions fixed. Let
\[
 \Delta F_l=F_h(\hat r_l(X))-F_h(R) = \mathcal K_h(\hat r_l(X) - r)v_h(\hat r_l(X) ) - \mathcal K_h(R-r)v_h(R). 
\]
Since $E_l[\epsilon_\Psi\mid X]=\rho_\Psi(X)$ and $E[\rho_\Psi(X)\mid R]=0$, iterated expectations give
\begin{align}\label{cf:indexmean}
 B_{\Psi,l}
 =E_l[\Delta F_l\epsilon_\Psi] =E_l[\{F_h(\hat r_l(X))-F_h(R)\}\rho_\Psi(X)] =E_l[F_h(\hat r_l(X))\rho_\Psi(X)].
\end{align}
For $\Psi=B$, \eqref{cf:rhoB} gives $\rho_B(X)=0$ almost surely, and hence $B_{B,l}=0$.

For $\Psi=A$, condition on the fitted index in \eqref{cf:indexmean} to obtain $B_{A,l}=E_l\!\left[  F_h(\hat r_l(X))\tau_A\{\hat r_l(X),\hat r_l\}\right]$.
Let $B_{A,l}^{\rm loc}$ denote the part of this expectation on $\{\hat r_l(X)\in\mathcal N_0\}$. Assumption~\ref{Aindex} and Lemma~\ref{cf:kernelbounds} imply
\begin{align*}
 \|B_{A,l}^{\rm loc}\|
 &\leq C_\rho a_{l,n}
 E_l[\|F_h(\hat r_l(X))\|]
 \leq C a_{l,n}h^{-d}E[\Omega(u_h(R))]
 =O_p(a_n).
\end{align*}

For a compactly supported kernel, $F_h(t)=0$ whenever $\|t-r\|_\infty>h$. For sufficiently small $h$, $\{t\in\mathbb R^d:\|t-r\|_\infty\leq h\} \subset\mathcal N_0$, so $B_{A,l}=B_{A,l}^{\rm loc}$.

For the Gaussian kernel, choose a fixed $\delta>0$ such that
$\{t\in\mathbb R^d:\|t-r\|_2<\delta\}\subset\mathcal N_0$.
Since $(1+s)e^{-s^2/2}\leq Ce^{-s^2/4}$ for all $s\geq0$,
the definition of $F_h$ gives
\begin{align*}
 \|F_h(t)\|_2
 &\leq Ch^{-d}
 \left(1+\frac{\|t-r\|_2}{h}\right)
 \exp\!\left(-\frac{\|t-r\|_2^2}{2h^2}\right)\leq Ch^{-d}
 \exp\!\left(-\frac{\|t-r\|_2^2}{4h^2}\right).
\end{align*}
For $t\notin\mathcal N_0$, we have $\|t-r\|_2\geq\delta$.
By conditional Jensen's inequality and iterated expectations,
\begin{align*}
 \|B_{A,l}-B_{A,l}^{\rm loc}\|_2
 &\leq E_l\!\left[
 \|F_h(\hat r_l(X))\|_2\,|\rho_A(X)|
 1\{\hat r_l(X)\notin\mathcal N_0\}
 \right]\leq Ch^{-d}
 \exp\!\left(-\frac{\delta^2}{4h^2}\right)
 E[|\rho_A(X)|].
\end{align*}
By conditional Jensen's inequality and Assumption~\ref{Areg}(iv),
$E[|\rho_A(X)|]\leq 2E[|A|]<\infty$.
Since $h=c_hn^{-\eta}$ with $\eta>0$, the last bound is
$o(\tau_n^{-1})$, uniformly over $l$.

Combining this tail bound with the local bound, and recalling that
the tail contribution is zero for a compactly supported kernel,
we obtain in both cases
\begin{equation}\label{cf:indexmeanrate}
 \max_{l\leq L}\|B_{A,l}\|_2
 =O_p(a_n)+o_p(\tau_n^{-1})
 =o_p(\tau_n^{-1}),
\end{equation}
where the last equality uses $\tau_n a_n\to0$.

Next, we bound \((P_l-E_l)[\Delta F_l\epsilon_\Psi]\), using its conditional second moment.
By Assumption~\ref{Areg}(iv) and
\eqref{cf:envelopemoment} with $T=\epsilon_\Psi^2$, $ h^{-d}E[\Omega(u_h(R))\epsilon_\Psi^2]=O(1)$.
On $\{a_{l,n}\leq h\}$, whose probability tends to one,
conditional independence within fold $l$, Lemma~\ref{cf:kernelbounds}(ii), and $\Omega^2\leq\Omega$ give
\begin{align*}%
 E_l\!\left[
 \|(P_l-E_l)(\Delta F_l\epsilon_\Psi)\|_2^2
 \right]
 &\leq \frac1{n_l}
 E_l[\|\Delta F_l\|_2^2\epsilon_\Psi^2]
 \leq
 \frac{C}{n_lh^{2d}}
 \left(\frac{a_{l,n}}h\right)^2
 E[\Omega(u_h(R))\epsilon_\Psi^2]
 \leq
 \frac{C}{n_lh^d}
 \left(\frac{a_{l,n}}h\right)^2
 =o_p(\tau_n^{-2}),
\end{align*}
where the last equality follows from
$\max_l a_{l,n}=O_p(a_n)$, $a_n/h\to0$, and
$n_l/n\to\omega_l>0$.
Conditional Chebyshev's inequality therefore yields
$ (P_l-E_l)(\Delta F_l\epsilon_\Psi)
 =o_p(\tau_n^{-1})$.

Combining it with \eqref{cf:indexmeanrate} (or $B_{B,l}=0$) and summing over the fixed number of folds gives
$ \sum_l n_l n^{-1}P_l[\Delta F_l\epsilon_\Psi]
 =o_p(\tau_n^{-1})$.
Adding the exactly pooled oracle average establishes \eqref{cf:residualtransfer}.

To prove \eqref{cf:taylortransfer}, write $e_l(X)=\hat r_l(X)-R$. 
On a sufficiently small fixed neighborhood $\mathcal N_1$ of $r$ whose closure lies in $\mathcal N$, Taylor's theorem yields
\begin{align*}
 q_{\Psi,l}(X)
 &=\gamma_\Psi(R)-v_h(\hat r_l(X))^\top\mathbf{c}_{\Psi,h}
 =\gamma_\Psi(R)-\gamma_\Psi(r)
   -\nabla\gamma_\Psi(r)^\top(R-r)
   -\nabla\gamma_\Psi(r)^\top e_l(X),\\
 |q_{\Psi,l}(X)|&\leq C\{\|R-r\|^2+a_{l,n}\}.
\end{align*}
By the polynomial-envelope bound of Lemma~\ref{cf:kernelbounds},
 $E_l[\|F_h(\hat r_l(X))\|\,|q_{\Psi,l}(X)|
                   1\{R\in\mathcal N_1\}]
 \leq C(h^2+a_{l,n})$.
Conditional Markov's inequality therefore bounds the corresponding fold average by $O_p(h^2+a_n)$. For a compact kernel there is no contribution from $R\notin\mathcal N_1$ for small $h$ on $a_{l,n}\leq h$, since a nonzero fitted weight implies $\|R-r\|_\infty\leq2h$. 

For the Gaussian kernel, choose a fixed $\delta>0$ such that
$\{t:\|t-r\|_2<\delta\}\subset\mathcal N_1$.
On $\{a_{l,n}\leq h\}$, for sufficiently small $h$,
\[
 |q_{\Psi,l}(X)|
 \leq C(1+\|u_h(R)\|_2)\{1+|\gamma_\Psi(R)|\},
 \qquad
 1\{R\notin\mathcal N_1\}
 \leq \frac{h^2}{\delta^2}\|u_h(R)\|_2^2.
\]
By Lemma~\ref{cf:kernelbounds}(ii) and
\eqref{cf:envelopemoment} with
$T=1+|\gamma_\Psi(R)|$ and $m=3$,
\begin{align*}
 &E_l\!\left[
 \|F_h(\hat r_l(X))\|_2\,|q_{\Psi,l}(X)|
 1\{R\notin\mathcal N_1\}
 \right] \leq
 Ch^{2-d}E\!\left[
 (1+\|u_h(R)\|_2)^3\Omega(u_h(R))
 \{1+|\gamma_\Psi(R)|\}
 \right]
 =O(h^2),
\end{align*}
where Assumption~\ref{Areg} gives local boundedness of
$\gamma_\Psi$ and
$E|\gamma_\Psi(R)|\leq E|\Psi|<\infty$.
Conditional Markov's inequality and $\tau_nh^2\to0$ yield
\begin{align*}
 &\left\|P_l\!\left[
 F_h(\hat r_l(X))q_{\Psi,l}(X)
 1\{R\notin\mathcal N_1\}
 \right]\right\|_2\leq
 P_l\!\left[
 \|F_h(\hat r_l(X))\|_2\,|q_{\Psi,l}(X)|
 1\{R\notin\mathcal N_1\}
 \right]
 =O_p(h^2)=o_p(\tau_n^{-1}).
\end{align*}

Finally, $\tau_n h^2\to0$ and $\tau_n a_n\to0$, followed by summation over fixed $L$, prove \eqref{cf:taylortransfer}.
\hfill $\square$

\section{Additional details for the simulation design}
\label{app:simcontinuity}

\paragraph{Population construction of $U$.}
For the covariate distribution in Section~\ref{sec:simulation},
let $\xi(X)=(X_1,X_2,X^\top\beta)^\top$ and define
\begin{align*}
 a_U&=\{E[\xi(X)\xi(X)^\top]\}^{-1}E[\xi(X)X_3],\qquad
 s_U^2=E[\{X_3-a_U^\top\xi(X)\}^2],
 \qquad U=\frac{X_3-a_U^\top\xi(X)}{s_U}.
\end{align*}
The quantities $a_U$ and $s_U$ are population constants
determined by the specified truncated covariate distribution.
We compute them once from the population formulas below
and use the same values in every Monte Carlo replication.
They are not estimated from the sample generated in
each replication.

Let $e_j$ denote the $j$th coordinate
vector in $\mathbb R^p$ and let $C_\xi = \big(e_1, e_2, \beta\big)^\top$ and 
 $G_\xi=C_\xi\Sigma C_\xi^\top$.
The symmetric truncation preserves $E[X]=0$.
Spherical symmetry of $\Sigma^{-1/2}X$ gives
\[
 \operatorname{cov}(X)=c_X\Sigma,\qquad
 c_X=\frac{E[Q_\chi\mid Q_\chi\leq b^2]}{p}
     =\frac{F_{\chi^2_{p+2}}(b^2)}{F_{\chi^2_p}(b^2)},
 \qquad Q_\chi\sim\chi^2_p,
\]
where $F_{\chi^2_k}$ denotes the chi-squared cdf with $k$ degrees
of freedom. Consequently,
\begin{align*}
 a_U&=G_\xi^{-1}C_\xi\Sigma e_3,\qquad 
 s_U^2=c_X\{\Sigma_{33}
       -e_3^\top\Sigma C_\xi^\top G_\xi^{-1}C_\xi\Sigma e_3\}.
\end{align*}
For $p=10$, $b=2.5$, $\Sigma_{jk}=0.5^{|j-k|}$, and
$\beta_j=0.5/j$, these formulas give
\begin{align*}
 c_X&\simeq0.470708994,\qquad
 a_U\simeq(-0.667226789,-0.015272176,1.334453578)^\top,\\
 s_U^2&\simeq0.224769112,\qquad s_U\simeq0.474098209.
\end{align*}
The displayed values are rounded. 
For this ellipsoidal truncation, the population linear-projection coefficients $a_U$ are unchanged,
while the variance of the population projection error is $c_X$ times its value under the untruncated $N(0,\Sigma)$ distribution.
 By construction, $E[U]=0$ and
$\operatorname{var}(U)=1$.

\paragraph{Population nuisance functions.}
Writing $t_z(x)=-0.3+x^\top\beta+(0.9+0.25x_2)z$,
the true nuisance functions are
\begin{align*}
 p(x)&=\Phi(x^\top\beta),\quad
 \pi(z,x)=\Phi(t_z(x)),\quad
 \mu(z,x)=x^\top\beta+
 \{x_1+\lambda U(x)\}\Phi(t_z(x))
 +0.5\phi(t_z(x)).
\end{align*}
Because $U$ is linear in $X$, the augmented outcome regressions
in Section~\ref{sec:simulation} contain $\mu(z,x)$ when evaluated
at the true propensity features. 
In the simulation, we estimate $\pi(z,x)$ by a separate probit regression of $D$ on an intercept and $X$ within the training observations with $Z=z$, for $z=0,1$.

\paragraph{IV validity and overlap
(Assumptions~\ref{ass:independence_cov}--\ref{ass:comsup_cov}
and~\ref{A1stCLATE}).}
Independence of $\upsilon$ from $(X,\varepsilon_1,\varepsilon_2)$
and the exclusion of $Z$ from the potential outcome equations give
conditional IV validity (Assumption~\ref{ass:independence_cov}).
Since $|X_2|\leq2.5$, the first-stage shift satisfies
$0.275\leq\gamma(X)\leq1.525$ throughout $\mathcal X$,
which gives conditional monotonicity and a strictly positive first
stage (Assumptions~\ref{ass:monotonicity_cov}
and~\ref{ass:first_stage_cov}). Also,
$ \Phi\!\left(-b\sqrt{\beta^\top\Sigma\beta}\right)  \leq p(X)\leq
\Phi\!\left(b\sqrt{\beta^\top\Sigma\beta}\right)$,
which gives $0.015\leq p(X)\leq0.985$.
This verifies common support across $Z$
(Assumption~\ref{ass:comsup_cov}) and the population overlap
restriction in Assumption~\ref{A1stCLATE} with $\varepsilon=0.01$.

Clipping $\hat p_l$ to $[0.01,0.99]$ enforces its fitted-overlap
restriction and bounds the inverse-probability weights by $100$.
Clipping $\hat\pi_l$ to $[10^{-4},1-10^{-4}]$ also satisfies the
range restriction in Assumption~\ref{A1stCLATE}.
The true propensity functions lie strictly inside their respective
clipping intervals, so uniform consistency makes clipping inactive
with probability approaching one.

\paragraph{Support, local regularity, and moments
(Assumptions~\ref{ass:support} and~\ref{Areg}).}
The vector $\xi(X)=C_\xi X$ has a smooth positive density on
$\{t:t^\top G_\xi^{-1}t<b^2\}$.
The map from $\xi(X)$ to $R$ is smooth and invertible, with
absolute Jacobian determinant $0.25\phi(t_0(X))\phi(t_1(X))>0$.
At $r=(\tilde x,\pi',\pi'')$, its inverse is
\[
 \xi_r=\left(\tilde x,
 \frac{\Phi^{-1}(\pi'')-\Phi^{-1}(\pi')-0.9}{0.25},
 0.3+\Phi^{-1}(\pi')\right)^\top.
\]
For the three evaluation points in Section~\ref{sec:simulation},
$\xi_r^\top G_\xi^{-1}\xi_r$ is approximately
$1.8664$, $0.6812$, and $0.9796$, respectively, all below $b^2=6.25$.
A sufficiently small neighborhood of each point therefore satisfies
the support and density conditions in Assumptions~\ref{ass:support}
and~\ref{Areg}(i).

Using $E[U\mid R]=0$, established below, we obtain
\begin{align*}
 \gamma_B(r)&=\pi''-\pi',\qquad
 \gamma_A(r)=\tilde x(\pi''-\pi')
 +\tfrac12\big[\phi\{\Phi^{-1}(\pi'')\}
              -\phi\{\Phi^{-1}(\pi')\}\big],
\end{align*}
which verifies the regression smoothness in Assumption~\ref{Areg}(ii).
The conditional first and second moments of $U$ are smooth at interior
values of $\xi(X)$, and the Gaussian threshold moments are smooth
in $t_z$, so the conditional covariance matrix of $(A,B)^\top$ is
continuous. 

Writing
$\varepsilon_2=\varepsilon_1/2+\sqrt{3}\,\zeta/2$, with
$\zeta\sim N(0,1)$ independent of $(X,\upsilon,\varepsilon_1)$,
shows that $A-\theta(r)B$ has a nonzero conditional Gaussian
innovation. Thus its conditional variance is positive, as required
by Assumption~\ref{Areg}(iii).
Moreover, $|U|\leq b/\sqrt{c_X}$, the covariates and true
inverse-probability weights are bounded, and the Gaussian errors
have all finite moments. Hence the local and global score-moment
conditions in Assumption~\ref{Areg}(iv) hold. Finally,
\[
 \sup_{x\in\mathcal X,\,z\in\{0,1\}}
 \operatorname{var}(Y\mid Z=z,X=x)
 \leq 2\bigl(b+|\lambda|b/\sqrt{c_X}\bigr)^2+8<\infty,
\]
which verifies the global conditional-variance bound in
Assumption~\ref{Areg}(v) required for the Gaussian kernel.

\paragraph{Kernel, bandwidth, and nuisance accuracy
(Assumptions~\ref{Akernel}, \ref{A1stCLATE}, and~\ref{Aindex}).}
The Gaussian kernel is allowed by Assumption~\ref{Akernel}.
The common bandwidth scale in Section~\ref{sec:simulation} is
$h\asymp n^{-6/35}$ with $d=3$, and
$1/7<6/35<1/3$, so its bandwidth restriction is satisfied.
Bounded support also gives the fourth moments needed for the
data-based bandwidth scales discussed in Section~\ref{sec:asymptotics}.
Under the usual fixed-dimensional estimation regularity, including
nonsingular probit information and outcome-regression design matrices,
the correctly specified regressions yield uniform root-$n$ nuisance
accuracy. All fitting uses the training sample, as required by
Assumption~\ref{A1stCLATE}. In particular,
\[
 \max_l(\mathcal E_{\mu,l}+\mathcal E_{\pi,l}+\mathcal E_{p,l})
 =O_p(n^{-1/2}),\qquad
 \tau_n\max_l\mathcal E_{p,l}(\mathcal E_{\mu,l}+\mathcal E_{\pi,l})
 =O_p(n^{-53/70})=o_p(1).
\]
These are the partial-$L_2$ and product-rate requirements
\eqref{cf:partialrates} in Assumption~\ref{A1stCLATE}.
The global squared nuisance errors are $O_p(n^{-1})$, so its
Gaussian-tail condition \eqref{cf:partialtail} holds with $C_0=0$.
For the index-accuracy part of Assumption~\ref{Aindex}, take
$a_n=n^{-1/2}$. Then
$a_n/h\asymp n^{-23/70}\to0$ and
 $\tau_na_n\asymp n^{-9/35}\to0$,
verifying \eqref{cf:indexaccuracy}.
Its projection-continuity and candidate-class requirements are
verified next, separately from these rate calculations.

\paragraph{Projection continuity and candidate-class membership
(Assumption~\ref{Aindex}).}
Write
 $\pi(z,x)=\Phi\big(a_z^0+(b_z^0)^\top x\big)$,
for  $z=0,1$, where 
 $(a_0^0,b_0^0)=(-0.3,\beta)$ and 
 $(a_1^0,b_1^0)=(0.6,\beta+0.25e_2)$.
Let $C_0=(e_1,b_0^0,b_1^0)^\top$
and $U = u^\top X$.
Since $\operatorname{cov}(X)$ is proportional to $\Sigma$,
the definition of $U$ implies
$u^\top\Sigma C_0^\top=0$
and $ \sup_{x\in\mathcal X}|u^\top x|
 =b\sqrt{u^\top\Sigma u}<\infty.$
For any full-row-rank matrix $C$, the spherical symmetry
of $\Sigma^{-1/2}X$ gives, by conditional reflection symmetry,
\[
 E[X\mid CX]
 =\Sigma C^\top(C\Sigma C^\top)^{-1}CX
 \qquad\text{almost surely}.
\]
Because $R$ and $C_0X$ generate the same sigma-field,
$E[U\mid R]=0$.
The conditional-mean formula above
therefore gives
\[
 \rho_A(X)=\lambda\Delta_\pi(X)U,
 \qquad \Delta_\pi(X)=\pi(1,X)-\pi(0,X).
\]

For the simulation design, choose the candidate classes
in Assumption~\ref{Aindex} as
$\mathcal R_n=\mathcal R(\delta)$ for every $n$, where
$\delta>0$ is a sufficiently small fixed constant and
\[
 \mathcal R(\delta)
 =
 \left\{
 q:\ 
 q(x)=
 \begin{pmatrix}
 x_1\\
 \Phi(a_0+b_0^\top x)\\
 \Phi(a_1+b_1^\top x)
 \end{pmatrix},
 \quad
 \max_{z\in\{0,1\}}
 \bigl(|a_z-a_z^0|+\|b_z-b_z^0\|_2\bigr)
 \leq\delta
 \right\}.
\]
This deterministic class does not depend on $n$.
We verify the projection-continuity condition
\eqref{cf:indexstability} and high-probability membership condition
\eqref{Emem} of Assumption~\ref{Aindex} for this same class.
For $q\in\mathcal R_n$, let
$C_q=(e_1,b_0,b_1)^\top$.
Since $C_0$ has rank three, $\delta$ can be chosen so that
$\sup_{q\in\mathcal R_n}
 \|(C_q\Sigma C_q^\top)^{-1}\|_{\mathrm{op}}<\infty$,
where $\|M\|_{\mathrm{op}}
=\sup_{\|v\|_2=1}\|Mv\|_2$ denotes the matrix
operator norm induced by the Euclidean norm.
All candidate and true probit indices lie in a fixed bounded
interval, on which $\phi$ is bounded away from zero.
The mean value theorem and
$\mathcal X=\{x:x^\top\Sigma^{-1}x\leq b^2\}$ give
\begin{align*}
 |a_z-a_z^0|
 +b\sqrt{(b_z-b_z^0)^\top\Sigma(b_z-b_z^0)}
 &=\sup_{x\in\mathcal X}
   |a_z-a_z^0+(b_z-b_z^0)^\top x|\\
 &\leq C\sup_{x\in\mathcal X}|q_z(x)-\pi(z,x)|
 \leq C\|q-r_0\|_\infty.
\end{align*}
Consequently,
$\|C_q-C_0\|_{\mathrm{op}}\leq C\|q-r_0\|_\infty$.
Since $q(X)$ and $C_qX$ generate the same sigma-field,
\begin{align*}
 |E[U\mid q(X)]|
 &=\left|
 u^\top\Sigma(C_q-C_0)^\top
 (C_q\Sigma C_q^\top)^{-1}C_qX
 \right|\\
 &\leq C\|C_q-C_0\|_{\mathrm{op}}
 \leq C\|q-r_0\|_\infty
 \quad\text{almost surely}.
\end{align*}
Let $\Delta_q(X)=q_1(X)-q_0(X)$.
By $|\Delta_q|\leq1$,
$|\Delta_\pi-\Delta_q|\leq2\|q-r_0\|_\infty$, and boundedness of $U$, 
\begin{align*}
 |E[\rho_A(X)\mid q(X)]|
 &\leq |\lambda|\,|\Delta_q(X)|\,|E[U\mid q(X)]|+|\lambda|E[|\Delta_\pi(X)-\Delta_q(X)|\,|U|\mid q(X)]\\
 &\leq C_\rho\|q-r_0\|_\infty\qquad\text{almost surely}.
\end{align*}
The constant $C_\rho$ is independent of $q$ and $n$.
This global bound verifies the local projection-continuity
condition \eqref{cf:indexstability} in Assumption~\ref{Aindex}.

For coefficient-consistent separate probit fits and fixed $L$,
$\Pr\!\big\{
 \max_{l\leq L,\,z\in\{0,1\}}
 \bigl(|\hat a_{l,z}-a_z^0|
       +\|\hat b_{l,z}-b_z^0\|_2\bigr)\leq\delta
 \big\}\longrightarrow1$,
which verifies the high-probability membership condition
\eqref{Emem} in Assumption~\ref{Aindex};
$r_0\in\mathcal R_n$ by construction.
The population treatment propensities satisfy
 $10^{-4}
 <\min_{z\in\{0,1\}}\inf_{x\in\mathcal X}\pi(z,x)$ and 
 $\max_{z\in\{0,1\}}\sup_{x\in\mathcal X}\pi(z,x)
 <1-10^{-4}$.
By compactness of $\mathcal X$ and continuity in the
probit coefficients, $\delta$ can be decreased so that
 $10^{-4}<q_z(x)<1-10^{-4}$ for all $q\in\mathcal R_n,\ x\in\mathcal X,\
 z\in\{0,1\}$.
Thus the clipping implemented in
Section~\ref{sec:simulation} is inactive on the
preceding coefficient-consistency event, and
the membership requirement \eqref{Emem} of
Assumption~\ref{Aindex} also holds for the implemented fitted indices.

\paragraph{Design diagnostics.}
Table~\ref{tab:simdesign} summarizes covariate explanatory power,
treatment-rate differences, and population propensity ranges.
The mean complier share $E[\pi(1,X)-\pi(0,X)]$ is approximately
$0.29$.
As a localization diagnostic, the kernel-weight effective sample
size $(\sum_i w_i)^2/\sum_i w_i^2$, with
$w_i=\mathcal K_h(R_i-r)$ evaluated at the true propensity
scores, ranges from $220$ to $390$ for $n=4000$ and from
$670$ to $1080$ for $n=20{,}000$, depending on $\tilde x$.
This diagnostic does not incorporate the local linear design
adjustment or conditional score variability.

\begin{table}[ht]
		\small
		\centering
		\caption{Explanatory power of the covariates and relevance of the instrument}
		\label{tab:simdesign}
		\begin{tabular}{lc}
			\toprule
			$R^2$ of $X$ for $Y$                                     & 0.28 \\
			McFadden pseudo $R^2$ of $X$ for $Z$                      & 0.13 \\
			McFadden pseudo $R^2$ of $(X,Z)$ for $D$                  & 0.27 \\
			\midrule
			First stage $E[D|Z=1]-E[D|Z=0]$                           & 0.46 \\
			Median complier share $\pi(1,X)-\pi(0,X)$                 & 0.30 \\
			\midrule
			Range of $p(X)$                                     & $[0.015,0.985]$ \\
			Range of $\pi(0,X)$                                 & $[0.007,0.968]$ \\
			Range of $\pi(1,X)$                                 & $[0.020,0.999]$ \\
			Range of $\gamma(X)$                                & $[0.275,1.525]$ \\
			\bottomrule
		\end{tabular}
		\par\smallskip
\begin{minipage}{\textwidth}\footnotesize\textit{Notes:}
Upper-panel statistics use an independent draw of 100,000
observations at $\lambda=0$.
The outcome-regression $R^2$ may change with $\lambda$;
the statistics involving only $D$, $Z$, and $X$ do not.
$R^2$ is from a linear regression, and the pseudo-$R^2$
values are McFadden measures from the corresponding logit
regressions.
Lower-panel endpoints are analytically determined population
extrema, rounded to three decimal places.
An affine index $a_0+a^\top X$ has extrema
$a_0\pm b\sqrt{a^\top\Sigma a}$ over $\mathcal X$.
\end{minipage}
\end{table}

	\end{appendix}
	
\end{document}